\documentclass[]{aa}
\usepackage{graphicx}
\usepackage[varg]{txfonts}
\usepackage{hyperref}
\newcommand{\x}{{\boldsymbol x}}
\newcommand{\s}{{\boldsymbol s}}
\newcommand{\D}{{\text{D}}}
\newcommand{\V}{{\text{V}}}
\DeclareMathOperator{\diam}{diam}
\DeclareMathOperator{\interior}{int}
\DeclareMathOperator{\spann}{span}
\begin{document} 
\title{Discontinuous Galerkin finite element methods for radiative transfer in spherical symmetry}
   \author{D. Kitzmann\inst{1}
          \and
          J. Bolte\inst{2}
          \and
          A. B. C. Patzer\inst{2}
          }
   \institute{Physikalisches Institut \& Center for Space and Habitability, University of Bern,
              Sidlerstr. 5, 3012 Bern, Switzerland\\
              \email{daniel.kitzmann@csh.unibe.ch}
         \and
             Zentrum f\"ur Astronomie und Astrophysik, Technische Universit\"at Berlin, 
             Hardenbergstr. 36, 10623 Berlin, Germany
             }

 \date{Received March 23, 2016; accepted July 24, 2016}

% \abstract{}{}{}{}{} 
% 5 {} token are mandatory
 
  \abstract{The discontinuous Galerkin finite element method (DG-FEM) is successfully applied to treat a broad variety of transport problems numerically. 
  	In this work, we use the full capacity of the DG-FEM to solve the radiative transfer equation in spherical symmetry. 
  	We present a discontinuous Galerkin method to directly solve the spherically-symmetric radiative transfer equation as a two-dimensional problem. 
  	The transport equation in spherical atmospheres is more complicated than in the plane-parallel case due to the appearance of an additional derivative with respect to the polar angle. 
  	The DG-FEM formalism allows for the exact integration of arbitrarily complex scattering phase functions, independent of the angular mesh 
resolution. 
  	We show that the discontinuous Galerkin method is able to describe accurately the radiative transfer in extended atmospheres and to capture 
discontinuities or complex scattering behaviour which might be present in the solution of certain radiative transfer tasks and can, therefore, cause 
severe numerical problems for other radiative transfer solution methods. 	
}


   \maketitle

\section{Introduction}

Many stars, such as AGB stars, have extended atmospheres which has important physical and observational implications. Especially the radiation 
field at large distances from the inner stellar disk becomes very dilute and is confined primarily to a narrow solid angle around the radial direction 
\citep{Mihalas1978}. Additionally, planets can also feature extended atmospheres in which case the use of the usual assumption of a 
plane-parallel atmosphere is no longer valid. This includes, for example, the extended and spherically stratified atmosphere of Pluto 
\citep{Gladstone2016Sci...351.8866G}.
Compared to the radiative transfer equation in the plane-parallel approximation, the transfer equation in spherical symmetry is 
more complicated and more difficult to handle. This is especially caused by the appearance of a derivative with respect to the polar angle, which is 
not present in the plane-parallel geometry. Other cases, in which the plane-parallel radiative transfer equation becomes much more 
complicated, include systems with a moving medium, where photons are subject to the Doppler effect and aberration \citep{Mihalas1978}.

Depending on the transport coefficients, the general radiative transport equation can have a very different mathematical character. As described in 
\citet{Kanschat2009nmmr.book.....K}, it is a hyperbolic wave equation in regions without matter, an elliptic diffusion equation in case of an 
optically thick medium, or a parabolic equation when the light is strongly peaked in the forward direction. It is, therefore, very difficult to find 
numerical methods, which are capable of dealing with these different (or even mixed) types of behaviour.

The problem of a grey spherical atmosphere in radiative equilibrium has been studied first by \citet{Kosirev1934MNRAS..94..430K} and 
\citet{Chandrasekhar1934MNRAS..94..444C}. An extension of the iterative moment method for plane-parallel atmospheres using variable Eddington factors 
\citep{Auer1970MNRAS.149...65A} to the problem of a spherically-symmetric atmosphere restricted to isotropic scattering has been introduced by 
\citet{Hummer1971MNRAS}. Additionally, approximative solutions based on a generalised Eddington approximation have been developed by 
\citet{Lucy1971ApJ...163...95L} and \citet{Unno1976PASJ...28..347U}, respectively.

Alternatively, the long \citep{Cannon1970ApJ...161..255C} or short characteristic methods \citep{Kunasz1988JQSRT..39...67K} can be used.
Combined with e.g. an operator perturbation method to account for the unknown source function -- such as the accelerated lamdba iteration 
\citep{Cannon1973JQSRT..13..627C} -- they can be 
used to solve the radiative transfer equation in multiple dimensions \citep{Olson1987JQSRT..38..325O, Kunasz1988JQSRT..39....1K}.
In combination with the work of \citet{Hummer1971MNRAS}, operator splitting methods have also been directly utilised for 
spherically-symmetric atmospheres by \citet{Kubat1994A&A...287..179K}.

Nowadays, the computationally demanding Monte Carlo techniques can solve very efficiently the radiative transfer problem for a broad variety of 
applications ranging from stellar atmospheres \citep[e.g.][]{1985ApJ...288..679A,1993ApJ...405..738L,1999A&A...344..282L}, circumstellar disks
\citep[e.g.][and the references therein]{2004A&A...417..793P}, supernovae \citep[e.g.][]{2005A&A...429...19L} to dusty objects \citep[e.g.][and the references therein]{2013ARA&A..51...63S}.
Their advantages are, for example, that they are intrinsically three-dimensional, can handle easily complex geometries and density distributions, and 
have a low algorithmic complexity. However, in some special cases with very optically thick regions the Monte Carlo approach becomes very slow because 
the number of photon interactions, i.e. scattering, increases exponentially. 
In these situations grid-based solution techniques, e.g. a finite difference, finite element, or finite volume method, for the differential equation of
radiative transfer are more favourable even if there are algorithmically quite complex. Moreover, grid-based techniques allowing for higher order 
approximations as well as error control and they can be formulated to be intrinsically flux conserving 
\citep{Hesthaven2008nodal, Henning2001IAUS..200..567H}.

In this study, we present a numerical method for a direct solution of the radiative transfer equation in spherical symmetry by means of a finite 
element method.
Finite element methods have already been previously used to solve the radiative transfer equation, most notably in the three-dimensional case by 
\citet{Kanschat1996PhDT........73K} or \citet{Richling2001A&A...380..776R}. 
% For the two-dimensional problem, a finite element method has been presented by \citet{Doicu2005JQSRT..91..347D}. 
In these cases, a continuous Galerkin finite element was used to discretise the transport operator. 
Such an approach, however, causes problems in cases where the transport equation has a dominant hyperbolic character, which may lead to 
discontinuities within the solution. It is possible to stabilise the numerical scheme by introducing a streamline diffusion method, for example. The 
streamline diffusion method essentially transforms the hyperbolic equation into a parabolic one by adding a small diffusion in the direction of photon 
transport. While this stabilises the numerical discretisation, it also adds a free parameter to the method, which has to be chosen carefully in such a 
way, that it stabilises the numerical scheme, but does not result in a decreased accuracy of the solution (see e.g. 
\citet{Eriksson1996cde..book.....E,Kanschat1996PhDT........73K}). An alternative way to deal with such problems is to soften the requirements for the 
solution of the finite element method, in particular the requirement of continuity of the solution across element boundaries. One such numerical 
approach in the context of finite element methods is the discontinuous Galerkin method.

The discontinuous Galerkin finite element method (DG-FEM) was first introduced by \citet{Reed_Hill_1973} to study neutron transport problems. Since 
then, the method has been successfully applied to a broad variety of parabolic, hyperbolic, and elliptic problems (e.g. \citet{Hesthaven2008nodal, 
Cockburn2003}). For three-dimensional internal heat transfer problems, the discretisation with a discontinuous Galerkin method has been studied by 
\citet{Cui2004NHTB...46..399C}. For astrophysical applications, discontinuous Galerkin methods have, for example, been used by 
\citet{Dykema1996ApJ...457..892D} and \citet{Castor1992ApJ...387..561C}.

In this work we use a DG-FEM to directly solve the radiative transfer equation with arbitrarily complex scattering 
phase functions in spherical symmetry as a two-dimensional problem. In contrast to other solution methods, the scattering integral can thereby be 
evaluated exactly, independent of the angular mesh resolution. The development of the numerical scheme is outlined for the formulation of the DG 
approach in Sects. \ref{sec:rt_eq} \& \ref{sec:dg_fem}. Its applicability to several test problems is presented in Sect. \ref{sec:test_problems}.
Remarks on the numerical efficiency are given in Sect. \ref{sec:efficiency}, followed by an outlook (Sect. \ref{sec:outlook}) and
a summary (Sect. \ref{sec:summary}).

\section{Radiative Transfer Equation}
\label{sec:rt_eq}

We consider the radiative transfer equation in spherical symmetry, cf. \citet{Mihalas1978}, i.e.

\begin{multline}
  \label{eq:rte_1}
    \mu \,\frac{\partial I_\nu(r,\mu)}{\partial r} + \frac{1-\mu^2}{r} \,\frac{\partial I_\nu(r,\mu)}{\partial \mu}
    = -\kappa_\nu(r)\, I_\nu(r,\mu) - \overline{s}_\nu(r) \,I_\nu(r,\mu)\\
      \shoveright{+ \eta_\nu^{\text{ind}}(r)\, I_\nu(r,\mu) + \eta_\nu^{\text{sp}}(r)}\\ 
      + \frac{1}{4\pi} \int_{-1}^{+1} \int_0^{2\pi} I_\nu(r,\mu') \,s_\nu(r,\mu, \mu'; \phi, \phi') \,d\mu' \, d\phi'
\end{multline}

with the specific intensity $I_\nu$, radius coordinate $r$, the cosine of the polar angle $\mu$, the azimuth angle $\phi$, absorption 
coefficient 
$\kappa_\nu$, total ($\overline{s}_\nu$) and differential ($s_\nu(r,\mu, \mu'; \phi, \phi')$) scattering coefficients, and induced and spontaneous 
emission coefficients $\eta_\nu^{\text{ind}}, \eta_\nu^{\text{sp}}$. 

The differential scattering coefficient can be decomposed in an averaged angle--independent part, the total scattering coefficient 
$\overline{s}_\nu(r)$, and the phase function $p(r,\mu,\mu';\phi,\phi')$. This
yields
\begin{equation}
  \begin{split}
    s_\nu & (r,\mu,\mu';\phi,\phi')\\ 
      & = \frac{1}{4\pi} \int_{-1}^{+1} \int_0^{2\pi} s_\nu(r,\mu,\mu';\phi,\phi')\,  d\mu' \,d\phi' \cdot p(r,\mu,\mu';\phi,\phi')\\ 
      & = \overline{s}_\nu(r) \cdot p(r,\mu,\mu';\phi,\phi').
  \end{split}
\end{equation}
The phase function describes the probability of photon scattering of incident directions $\mu'$ and $\phi'$ into the directions $\mu$ and $\phi$. 
Since in spherical symmetry the radiation field is independent of the azimuth angle $\phi$, the phase function $p(r,\mu,\mu';\phi,\phi')$ can formally 
be integrated with respect to the azimuth angle to obtain the azimuthally averaged phase function
\begin{equation}
  p^{(0)} (\mu',\mu) := \frac{1}{2\pi} \int_0^{2\pi} p(\mu',\mu;\phi,\phi')\, d\phi',
\end{equation}
which is then used in the transfer equation.
To guarantee the conservation of energy, the phase function is normalised to unity, i.e.
\begin{equation}
  \frac{1}{2} \int_{-1}^{+1} p^{(0)}(\mu,\mu') \, d\mu' = 1.
\end{equation}

\begin{figure}
  \centering
  \includegraphics[scale=0.65]{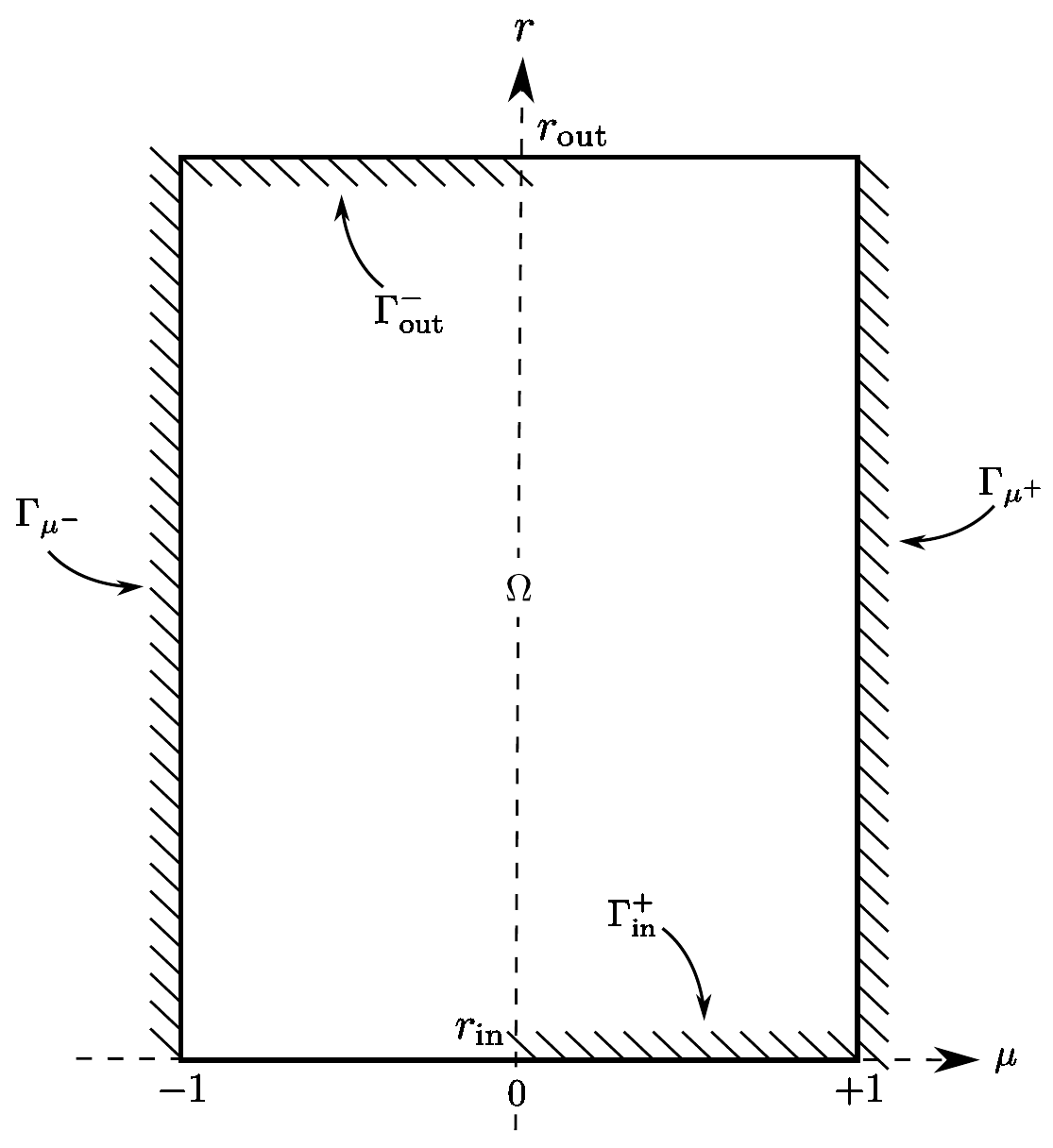}
  \caption{Illustration of the computational domain $\Omega$. The boundaries of the domain between the inner radius 
$r_\mathrm{in}$ and outer radius $r_\mathrm{out}$, as well as $\mu$ between $+1$ and $-1$, are denoted by the respective $\Gamma$.}
  \label{fig:domain}
\end{figure}

The extinction coefficient $\chi_\nu = \kappa_\nu + \overline{s}_\nu$ can be combined with the induced emission into
\begin{equation}
  \hat{\chi}_\nu = \chi_\nu - \eta_\nu^{\text{ind}} \, .
\end{equation}
In this work we only consider cases with $\chi_\nu > \eta_\nu^{\text{ind}}$ and therefore $\hat{\chi}_\nu > 0$, i.e. we do not consider systems which 
are dominated by induced emission (such as maser). Equation~\eqref{eq:rte_1} is therefore simplified to
\begin{equation}
  \label{eq:rte_2}
  \mu\, \frac{\partial I_\nu}{\partial r} + \frac{1-\mu^2}{r} \,\frac{\partial I_\nu}{\partial \mu} = -\hat{\chi}_\nu \,I_\nu + \eta_\nu^{\text{sp}} 
        +  \frac{\overline{s}_\nu}{2} \int_{-1}^{+1} I_\nu(\mu') \,p^{(0)}(\mu,\mu') \,d\mu',
\end{equation}
where all direct dependences on $r$ and $\mu$ have been omitted to simplify the presentation.
Instead of the usual form of the radiative transfer equation in spherical symmetry given by Eq.~\eqref{eq:rte_2} we will use the mathematically 
equivalent form

\begin{multline}
  \label{eq:rte_3}
    \frac{\partial}{\partial r}\left( \mu \,I_\nu \right) + \frac{\partial}{\partial \mu} \left( \frac{1-\mu^2}{r}\, I_\nu \right)
    + \frac{2\,\mu}{r} I_\nu\\ 
     =  -\hat{\chi}_\nu \,I_\nu + \eta_\nu^{\text{sp}}
     + \frac{\overline{s}_\nu}{2} \int_{-1}^{+1} I_\nu(\mu') \,p^{(0)}(\mu,\mu') \,d\mu'.
\end{multline}

To simplify the notation we drop the explicit frequency dependence in the transfer equation and its corresponding coefficients in the following.
Note, that we do not introduce the usual (radial) optical depth in the transfer equation. Doing so -- by moving the extinction coefficient 
$\hat{\chi}$ to the left hand side -- would introduce additional terms of the form $1/\hat{\chi}$ which can be numerically destabilising since the 
extinction coefficient can vary over several orders of magnitude.

The Eq.~\eqref{eq:rte_3} is solved on the two-dimensional domain $\Omega \subset \mathbb{R}^2$ 
with $\Omega \ni \x = (r,\mu) \in (r_\text{in},r_\text{out})\times (-1,1)$ as displayed in Fig.~\eqref{fig:domain}. The inner (outer) radius of $\Omega$
is given by $r_\text{in}$ ($r_\text{out}$).

At the inner boundary $\Gamma_{\text{in}}$ of the domain boundary conditions need to be imposed for positive values of $\mu$, i.e.
\begin{equation}
  I(\x) = I_{\text{in}}\quad  \text{for } \x \in \Gamma_{\text{in}}^+
\end{equation}
given by the radiation field of e.g. the stellar interior. Likewise at the outer boundary $\Gamma_{\text{out}}$ an incident external radiation field 
has to be prescribed for negative values of $\mu$, i.e.
\begin{equation}
  I(\x) = I_{\text{out}}\quad  \text{for } \x \in \Gamma_{\text{out}}^-.
\end{equation}
Note, that since the differential with respect to $\mu$ in Eq.~\eqref{eq:rte_3} vanishes on the boundary $\partial \Gamma$, no explicit boundary 
conditions need to be imposed on $\Gamma_{\mu^\pm}$.
Therefore, the boundary value problem considered in this work is given by
\begin{equation}\label{eq:rt_problem}
\begin{aligned}
\frac{\partial}{\partial r}&\left( \mu \, I \right) + \frac{\partial}{\partial \mu} \left( \frac{1-\mu^2}{r} \, I \right) \\
      &= - \frac{2\, \mu}{r} \, I -\hat{\chi}\,  I + \eta^{\text{sp}} 
      + \frac{\overline{s}}{2} \int_{-1}^{+1} I(\mu') \, p^{(0)}(\mu,\mu') \, d\mu'&& \text{in } \Omega,\\
\frac{\partial I}{\partial r} & =
      -\frac{2}{r} \, I \mp \hat{\chi}\,  I \pm \eta^{\text{sp}} 
      \pm \frac{\overline{s}}{2} \int_{-1}^{+1} I(\mu') \, p^{(0)}(\pm 1,\mu') \, d\mu'&& \text{on } \Gamma_{\mu^\pm},\\
I & = I_{\text{in}} &&\text{on } \Gamma_{\text{in}}^+,\\
I & = I_{\text{out}} &&\text{on } \Gamma_{\text{out}}^-.
\end{aligned}
\end{equation}

\section{Formulation of the discontinuous Galerkin method}
\label{sec:dg_fem}

\begin{figure}[htb]
  \centering
  \includegraphics[scale=0.60]{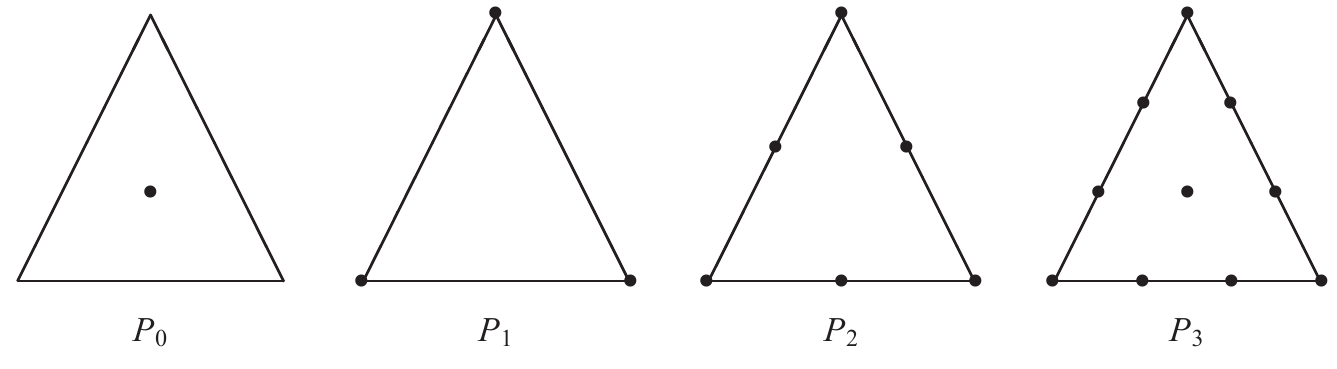}\\
  \includegraphics[scale=0.60]{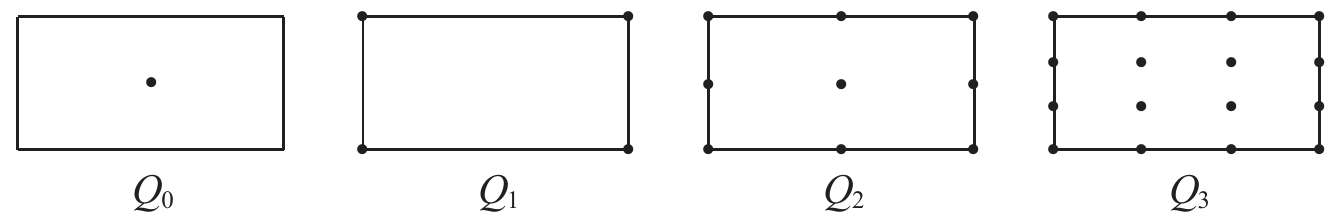}
  \caption{Lagrangian reference elements for dimension $n=2$, and polynomial degrees $q=0,1,2,3$. The degrees of freedom (see Table \ref{tbl:1}) are 
marked with $\bullet$. \textit{Top}: $P_q$ elements, \textit{bottom}: $Q_q$ elements.}
  \label{fig:P_q_Q_q}
\end{figure}

For the formulation of the discontinuous Galerkin method we are following the notation of \citet{Hesthaven2008nodal}.
The domain $\Omega$ is described by $K$ non--overlapping elements $\D^k$, i.e.
\begin{equation}
 \interior(\D^k) \cap \interior(\D^n) = \emptyset \quad \text{for } k\neq n,
\end{equation}
such that the continuous domain $\Omega$ is approximated by the finite domain $\Omega_h$ with
\begin{equation}
  \Omega \backsimeq \Omega_h = \bigcup_{k=1}^{K} \D^k = \overline{\Omega}
\end{equation}
with a local element size $h^k:=\diam(\D^k)$.
\begin{definition}
For any open set $\Omega$ in $\mathbb{R}^n$ let $P_q(\Omega)$ and $Q_q(\Omega)$ denote the
set of algebraic polynomials of total and partial degree $\le q$, respectively. For $n=2$
Table~\ref{tbl:1} shows the monomial basis functions of $P_q$ and $Q_q$ for $q=0,1,2,3$
and Fig.~\ref{fig:P_q_Q_q} displays their Lagrangian elements.
\end{definition}

\begin{table}[h]
\caption{Monomial basis of polynomial spaces $P_q$ and
$Q_q$ of total and partial degree $\leq q$, respectively, for $n=2$, and $q=0,1,2,3$.}\label{tbl:1}
\centering
\begin{tabular}{cc}
 & polynomial in $x,y$ \\
$q$& $P_q$\\ \hline
0 & 1\\
1 & \qquad\qquad\qquad$1,x,y$\qquad\qquad\qquad\qquad\\
2 & $1,x,y,x^2,xy,y^2$\\
3 & $1,x,y,x^2,xy,y^2,x^3,x^2y,xy^2,y^3$\\
 & \\
 & $Q_q$ \\ \hline 
0 & 1\\
1 & $1,x,y,xy$ \\
2 & $1,x,y,x^2,xy,y^2,x^2y,x^2y^2,xy^2$\\
3 & $1,x,y,x^2,xy,y^2,x^2y,x^2y^2,xy^2,$\\
  &$x^3,y^3,x^3y,x^3y^2,x^3y^3,x^2y^3,xy^3$
\end{tabular}
\end{table}

In this work we use the nodal representation of the solution $I_h^k:=I_h|_{\D^k}$ on each element. For each element $\D^k$ a suitable set of local 
grid points $\x_i^k \in \D^k$ is chosen and the local continuous solution of $I_h^k$ is expressed using the interpolating Lagrange polynomials, such 
that
\begin{equation}
I_h^k(\x) := \sum_{i=1}^{N_p} I_h^k(\x_i^k) \,\ell_i^k(\x)\quad \text{for all } \x \in \D^k
\label{eq:sol_exp_local}
\end{equation}
is a polynomial of order $q$ on each element. Here the so called trial functions $\ell_i^k$ are
the interpolating two--dimensional Lagrange polynomials through the grid points $\x_i^k$, i.e.
\begin{equation}
\ell_i^k(\x) :=\ell_i^k(x) \,\ell_i^k(y) = \prod_{\substack{j=1\\j\neq i}}^{N_p} \frac{(x-x_j)}{(x_i-x_j)} \, \prod_{\substack{j=1\\j\neq i}}^{N_p} 
\frac{(y-y_j)}{(y_i-y_j)} 
\end{equation}
For a given order $q$ the number of local grid points $N_p$ is given by
\begin{equation}
  N_q = \frac{(q+1)\,(q+2)}{2}
\end{equation}
for $P_q$ elements and by
\begin{equation}
  N_q = (q+1)^2
\end{equation}
for $Q_q$ elements, respectively. In principle the polynomial order $q$ and thus $N_p$ can vary from element to element. However, for simplicity of 
notation we assume here that all elements have the same order $q$. 
Subsequently the set $\mathcal N^k$ of the local grid point indices within each element $k$ is denoted by
\begin{equation}
  \mathcal N^k = \lbrace 1, \dotsc, N_p  \rbrace.
  \label{eq:set_local_ind} 
\end{equation}
The global piecewise continuous solution can then be written as the direct sum of all $K$ solutions
\begin{equation}
  I(\x) \backsimeq I_h(\x) = \bigoplus_{k=1}^{K} I_h^k(\x).
  \label{eq:sol_exp_glob}
\end{equation}
The trial functions $\ell_i^k(\x)$ introduced in Eq.~\eqref{eq:sol_exp_local} form a basis of the finite vector space $\V_h$. This vector space is 
locally defined by $\V_h^k = \spann\lbrace \ell_i^k(\D^k) \rbrace_{i=1}^{N_p}$ and globally given by $\V_h=\bigoplus_{k=1}^{K} \V_h^k$. Note, that 
unlike the case of the continuous Galerkin finite element method \citep{Eriksson1996cde..book.....E} no continuity of the trial functions (and 
therefore also of the solution $I$) is assumed or required across
inter--element boundaries.

We define the residual of Eq.~\eqref{eq:rt_problem} by
\begin{equation}
\begin{split}
  \mathcal R_h (I_h(\x)) = &\frac{\partial}{\partial r}\left( \mu\, I_h \right) + \frac{\partial}{\partial \mu} \left( \frac{1-\mu^2}{r}\,  I_h 
\right)\\ 
                                    &+ \frac{2\, \mu}{r}\,  I_h +\hat{\chi} \, I_h - \eta^{\text{sp}} - \frac{\overline{s}}{2} \int_{-1}^{+1} 
I_h(\mu') \, p^{(0)}(\mu,\mu') \, d\mu'
\end{split}
\end{equation}
with $I_h \in \V_h$.
In the context of finite element methods this residual is minimised with respect to all test functions $\phi_h^k \in C_\text{c}^\infty(\D^k)$ in each 
element. This gives the weak formulation
of Eq.~\eqref{eq:rt_problem}, which represents the physical principle of virtual work. Thus it is required that
\begin{equation}
  \int_{\D^k} \mathcal R_h(\x) \,\phi_h^k(\x) \, d\x = 0 \quad \text{for all } \phi_h^k \in C_\text{c}^\infty(\D^k) \ \text{and } k=1,\dotsc,K.
\end{equation}

For this study we consider only test functions, which are members of the same finite vector space as the trial functions, i.e. $\phi^k_h \in \V^k_h 
\subset C_\text{c}^\infty(\D^k)$. This corresponds to a
Ritz--Galerkin method \citep{Hesthaven2008nodal}. We note, that in principle $\phi^k_h$ may also be defined as a member of vector spaces other than 
$\V^k_h$ (e.g. $\delta$-distributions). The choice of
different trial and test vector spaces leads to Petrov--Galerkin finite element schemes \citep{Hesthaven2008nodal} which are, however, not considered 
here.

For this particular choice of test functions we require for each element $\D^k$
\begin{equation}
  \label{eq:element_res1}
  \int_{\D^k} \mathcal R_h(\x)\, \phi_h^k(\x)\,  d\x = 0 \quad \text{for all } \phi_h^k \in \V^k_h \ \text{and } k=1,\dotsc,K,
\end{equation}
i.e. the residual is required to be orthogonal to all test functions $\phi_h^k \in \V_h^k$.
Since the test functions are given by a linear combination of the Lagrange polynomials, i.e.
\begin{equation}
  \phi_h^k = \sum_{i=1}^{N_p} \hat{\phi}_i^k \, \ell_i^k (\x)
\end{equation}
with the expansion coefficients $\hat{\phi}_h^k \in \mathbb{R}$, Eq.~\eqref{eq:element_res1} results in the $N_p$ equations
\begin{equation}
  \label{eq:element_res2}
  \int_{\D^k} \mathcal R_h(\x)\, \ell_i(\x) \, d\x = 0 \quad \text{for } i=1,\dotsc,N_p
\end{equation}
on each element $\D^k$.

To sum up, we consider in this work the following DG($q$)-method:\\
Seek $I_h \in \V_h$ such that
\begin{equation}\label{eq:dg_method}
\begin{aligned}
\int_{\D^k} & \bigg\{ \frac{\partial}{\partial r}\left( \mu \, I_h^k \right) + \frac{\partial}{\partial \mu} \left( \frac{1-\mu^2}{r} \, I_h^k 
\right) + \frac{2\, \mu}{r} \, I_h^k + \hat{\chi}\,  I_h^k - \eta^{\text{sp}} \\
      & - \frac{\overline{s}}{2} \int_{-1}^{+1} I_h^k(\mu') \, p^{(0)}(\mu,\mu') \, d\mu' \bigg\} \,\ell_i^k(\x)\, d\x = 0 && \text{in } \Omega,\\
\frac{\partial I_h}{\partial r}  =
      &-\frac{2}{r} \, I_h \mp \hat{\chi}\,  I_h \pm \eta^{\text{sp}} \\
      &\pm \frac{\overline{s}}{2} \int_{-1}^{+1} I_h(\mu') \, p^{(0)}(\pm 1,\mu') \, d\mu'&& \text{on } \Gamma_{\mu^\pm},\\
I_h = & I_{\text{in}} &&\text{on } \Gamma_{\text{in}}^+,\\
I_h = & I_{\text{out}} &&\text{on } \Gamma_{\text{out}}^-
\end{aligned}
\end{equation}
for $i=1,\dotsc,N_p$ and $k=1,\dotsc,K$.

\subsection{Elementwise calculations}
\label{sec:elment_calculations}

Let us first consider the two terms $\frac{2\,\mu}{r}\, I_h^k +\hat{\chi}\, I_h^k$ under the integral in Eq.~\eqref{eq:dg_method}. 
Using the expansion of the solution $I_h^k$ within one element $k$ (see Eq.~\eqref{eq:sol_exp_local}) yields
\begin{equation}
  \begin{split}
    \int_{\D^k} \left( \frac{2\,\mu}{r} + \hat{\chi} \right) \, I_h^k \, \ell_i^k \,d\x = \sum_{j=1}^{N_p} I_h^k (\x_j^k) \int_{\D^k} & \left( 
    \frac{2\,\mu}{r}  + \hat{\chi} \right) \,\ell^k_j
    \,\ell^k_i \,d\x \\ 
    & \text{for all } i,j \in \mathcal N^k.
  \end{split}
\end{equation}
This defines the local symmetric mass matrix $\mathcal M^k=(\mathcal M_{ij}^k)$ with
\begin{equation}
  \mathcal M_{ij}^k := \int_{\D^k} \left( \frac{2\, \mu}{r} + \hat{\chi} \right)\,  \ell^k_j \, \ell^k_i d\x \quad \text{for all } i,j \in \mathcal 
N^k.
  \label{eq:mass_matrix}
\end{equation}
The term $\eta^{\text{sp}}$ in Eq.~\eqref{eq:dg_method} simply yields the local load vector $\mathcal B^k=(\mathcal B_{i}^k)$, with
\begin{equation}\label{eq:local_load_vector}
  \mathcal B^k_i := \int_{\D^k} \eta^{\text{sp}} \,\ell_i^k d\x \quad \text{for all } i \in \mathcal N^k.
\end{equation}

Next we consider the differential with respect to $r$ in Eq.~\eqref{eq:dg_method}. An integration by parts shifts the differential from the unknown 
quantity $I_h^k$ to the test function $\ell^k_i$ introducing an additional integral over the boundary of the element $\D^k$, i.e.
\begin{equation}
  \begin{split}
    \int_{\D^k} \frac{\partial}{\partial r} \left( \mu \,I_h^k \right) \,\ell^k_i d\x = - \int_{\D^k} \mu \, I_h^k \, \frac{\partial 
\ell^k_i}{\partial r} d\x + \int_{\partial \D^k} & \hat{\boldsymbol n}_r \cdot \mu \, I_h \ell^k_i d\s \\ 
  & \text{for all } i \in \mathcal N^k \ ,
  \end{split}
\end{equation}
where $\hat{\boldsymbol n}_r$ denotes the outward facing normal vector in $r$ direction (see Fig.~\ref{fig:jump}).

Since by being allowed to be discontinuous across element boundaries, the solution $I_h$ is not uniquely defined at the boundary. Therefore, a so 
called numerical flux $(f)^*$ is introduced which describes the flow of information from one element to another, yielding
\begin{equation}
  \int_{\partial \D^k} \hat{\boldsymbol n}_r \cdot \mu\, I_h \, \ell^k_i d\s = \int_{\partial \D^k} \hat{\boldsymbol n}_r \cdot (f)^*(I_h^k,I_h^n) \, 
\ell_i^k d\s.
\end{equation}
The numerical flux will in general depend on the solutions of the local element $I_h^k$ and the adjacent element $I_h^n$ and is directly related to 
the dynamics of the considered differential equation. For a particular choice of the flux a matrix $\mathcal F^{kn}$ can be defined which contains the 
contributions due to the jump at the inter--element boundaries between the elements $k$ and $n$. The exact contribution depends on the details of the 
numerical flux used, as discussed in Sect.~\ref{sec:flux}.

Using again the expansion of Eq.~\eqref{eq:sol_exp_local} for the solution in the element $k$, the local stiffness matrix $\mathcal S^k_{r}=(\mathcal S_{ij,r}^k)$ can be 
introduced as
\begin{equation}
  \mathcal S^k_{ij,r} := - \int_{\D^k} \mu\, \ell^k_j \,\frac{\partial \ell^k_i}{\partial r} \, d\x \quad \text{for all } i,j \in \mathcal N^k.
\label{eq:stiffness_matrix_r}
\end{equation}
For the differential with respect to $\mu$ in Eq.~\eqref{eq:dg_method} integration by parts yields

\begin{equation}
  \begin{split}
  \int_{\D^k} \frac{\partial}{\partial \mu} \, &\left(\frac{1-\mu^2}{r} \, I_h^k \right) \, \ell^k_i \,d\x\\ =& - \int_{\D^k} \frac{1-\mu^2}{r} \, I_h^k 
\frac{\partial \ell^k_i}{\partial \mu} \, d\x
 + \int_{\partial \D^k} \hat{\boldsymbol n}_\mu \cdot \frac{1-\mu^2}{r} \, I_h \ell^k_i \,d\s\\
                  =& - \int_{\D^k} \frac{1-\mu^2}{r}\, I_h^k  \,\frac{\partial \ell^k_i}{\partial \mu} \,d\x 
                   + \int_{\partial \D^k} \hat{\boldsymbol n}_\mu \cdot (f)^* \ell^k_i \,d\s
  \end{split}
\end{equation}

for all $i \in \mathcal N^k$ and the respective outward facing normal vector in $\mu$ direction (see Fig.~\ref{fig:jump}).

The corresponding stiffness matrix is given by
\begin{equation}
  \mathcal S^k_{ij,\mu} := - \int_{\D^k} \frac{1-\mu^2}{r} \, \ell_j^k \frac{\partial \ell_i^k}{\partial \mu}\, d\x \quad \text{for all } i,j \in 
\mathcal N^k.
\label{eq:stiffness_matrix_mu}
\end{equation}
The total local stiffness matrix $S^k=(\mathcal S_{ji}^k)$ for the element $k$ can then be obtained by the sum of $S^k_{ij,\mu}$ and $S^k_{ij,r}$, i.e.
\begin{equation}
  \mathcal S^k_{ij} := \mathcal S^k_{ij,r} + \mathcal S^k_{ij,\mu}.
\end{equation}

\subsection{Numerical flux}\label{sec:flux}
\begin{figure}
  \centering
  \includegraphics[scale=1.1]{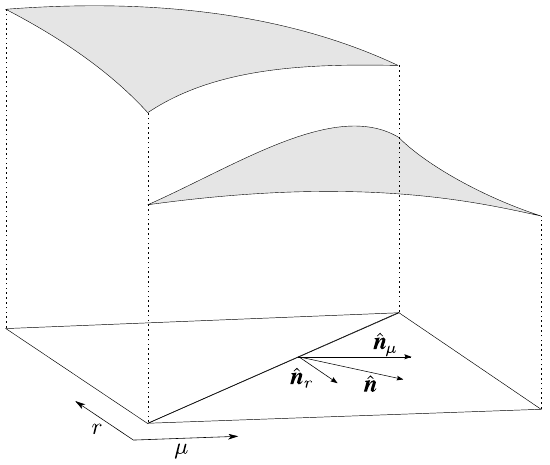}
  \caption{Illustration of the jump in the numerical solution along the normal $\hat{\boldsymbol n}$ between $P_2$ elements. The normals in the 
directions of $r$ and $\mu$ are denoted by $\hat{\boldsymbol n}_r$ and $\hat{\boldsymbol n}_\mu$, respectively.}
  \label{fig:jump}
\end{figure}

The numerical flux considered here is the so called upwind flux (e.g. \citet{Cockburn2003,Hesthaven2008nodal}), given by
\begin{equation}
  (f)^*=(a I)^* = \lbrace \lbrace a I \rbrace \rbrace + \tfrac{1}{2} \left| a \right| \llbracket I \rrbracket
\end{equation}
with
\begin{equation}
  \lbrace \lbrace I \rbrace \rbrace = \frac{I^- + I^+}{2}
\end{equation}
as the average of the solutions and
\begin{equation}
  \llbracket I \rrbracket = \hat{\boldsymbol n}^- I^- + \hat{\boldsymbol n}^+ I^+
\end{equation}
as the jump of the solution along the normal. 
The situation is shown in Fig.~\ref{fig:jump}, where the jump between two $P_2$ elements is illustrated. The parameter $a$ is here either given by 
\begin{equation}
  a = \mu
\end{equation}
for the differential with respect to $r$, or by
\begin{equation}
  a = \frac{1-\mu^2}{r}
\end{equation}
for the differential with respect to $\mu$.
The particular choice of this flux can be motivated by the fact, that for the radiative transfer problem considered here information propagates only 
in the direction of photon travel. The upwind flux for radiative transfer problems has already been considered by e.g. \citet{Li2006}. Note, however, 
that the choice of this flux is not unique. Other fluxes with different properties may be used (see \citet{Cockburn2003} for a review on DG fluxes).
The connection between the test and trial functions of the adjacent elements can be written as a matrix
\begin{equation}
  \begin{split}
    \mathcal F^{kn}_{ij} := \int_{\partial \D^k} \hat{\boldsymbol n}_r \cdot (f)^*\!\left(I_h^k(\ell_i^k),I_h^n(\ell_j^n)\right)& \, \ell_i^k d\x \\ 
    & \text{for all } i \in \mathcal N^k \text{ and } j \in \mathcal N^n.
  \end{split}
  \label{eq:dg_jump}
\end{equation}
Unlike the other matrices (such as $\mathcal M^k$), which are only defined over a single element, the matrix $\mathcal F^{kn}$ connects a local 
element $k$ with adjacent elements $n$. Not taking into account the scattering integral considered below, this would be the only connection between 
different elements.

\subsection{Scattering matrix}
The contribution due to scattering in Eq.~\eqref{eq:dg_method} is given by the integral
\begin{equation}
  \int_{\D^k} \left(\frac{\overline{s}}{2} \int_{-1}^{+1} I_h^k(r,\mu') \,p^{(0)}(r,\mu,\mu')\, d\mu' \right)\, \ell_i^k(\x) \, d\x \quad \text{for 
all } i \in \mathcal N^k.
\end{equation}
Using the expansions from Eqs.~\eqref{eq:sol_exp_local} \& \eqref{eq:sol_exp_glob} results in
\begin{multline}
    \int_{\D^k} \left(\frac{\overline{s}}{2} \int_{-1}^{+1} I_h^k(r,\mu') \, p^{(0)}(r,\mu,\mu')\,  d\mu' \right) \, \ell_i^k(\x)\,  d\x\\
   = \sum_{n=1}^{K} \sum_{j=1}^{N_p} I_h^k(r_j^n,\mu_j^n)\\
   \times \int_{\D^k} \left( \overline{s}(r) \, \int_{-1}^{+1} \ell_j^n(r,\mu') \, 
p^{(0)}(r,\mu,\mu') \, d\mu' \right) \, \ell_i^k(\x) \, d\x
\end{multline}
for all $i \in \mathcal N^k$ and $\ j \in \mathcal N^n$.
This defines a matrix $\mathcal J^{kn}$ containing the contributions due to scattering
\begin{multline}
   \mathcal J^{kn}_{ij} = \int_{\D^k} \left( \frac{\overline{s}}{2} \int_{-1}^{+1} \ell_j^n(r,\mu') \, p^{(0)}(r,\mu,\mu')\,  d\mu' \right) \, \ell_i^k(\x)\,  d\x\\
 \quad \text{for all } i \in \mathcal N^k \text{ and } j \in \mathcal N^n.
 \label{eq:scattering_integral}
\end{multline}
Note, that the scattering integral in Eq.~\eqref{eq:scattering_integral} only contains known quantities, namely the scattering phase function 
$p^{(0)}$ and the Lagrange polynomials $\ell_j$. Thus, the integral can be evaluated exactly for any arbitrary complex phase function, independent of 
the mesh resolution. This is an direct advantage over other radiative transfer methods, e.g. the discrete ordinate methods 
\citep{Thomas2002rtao.book.....T}, where the evaluation of the scattering integral is directly coupled to the resolution of the angular mesh 
\citep{Chandrasekhar1960ratr.book.....C}.

\subsection{Transformation matrix}
\label{sec:transformation_matrix}

To assemble a global matrix for the solution of the resulting system of equations, the matrixes and tensors with the local indices $i$ and $j$ must be 
mapped onto a matrix with global indices. This is done via a transformation matrix $\mathcal T$.
Let $g$ be a linear bijective mapping, which maps the set $\mathcal N^k$ of local indices $i$ within each element $k$ onto a set $\mathcal N$ of 
global coordinates
\begin{equation}
  \mathcal N = \lbrace 1, ..., k \cdot N_p  \rbrace
\end{equation}
with
\begin{equation}
  g: \mathcal N^k \rightarrow \mathcal N , \quad g(i,k) = \alpha \in \mathcal N.
\end{equation}
The linear mapping can be written as a transformation matrix $\mathcal T^k$. The matrix $\mathcal T_{\alpha i}^{k}$ then maps, for example, the 
scattering matrix $\mathcal J^{kn}$ onto the global scattering matrix $\mathcal J$ with the global indices $\alpha$ and $\beta$, which yields
\begin{equation}
  \mathcal J_{\alpha \beta} = \sum_{k=1}^{K} \sum_{n=1}^{K} \sum_{i=1}^{N_p} \sum_{j=1}^{N_p} \mathcal T_{\beta j}^{n}\, \mathcal T_{\alpha i}^{k} \, \mathcal J^{kn}_{ij}
\end{equation}
Likewise the global mass matrix $\mathcal M$, stiffness matrix $\mathcal S$, and jump matrix $\mathcal F$ can be obtained. The total left hand side 
matrix $\mathcal A$ of the linear system of equations can be expressed by:

\begin{equation}
  \mathcal A_{\alpha \beta} = \mathcal J_{\alpha \beta} + \mathcal M_{\alpha \beta} + \mathcal S_{\alpha \beta} + \mathcal F_{\alpha \beta} \, .
\end{equation}
The resulting linear system of equations is then
\begin{equation}
  \mathcal A \cdot \mathbf I = \mathcal B
	\label{eq:problem_lse}
\end{equation}
with $\mathbf I_\alpha := \sum_{k=1}^{K}  \sum_{i=1}^{N_p} \mathcal T_{\alpha i}^{k} \, I_h^k(\x_i^k)$ and $\mathcal B$ the global load vector (cf. Eq. \eqref{eq:local_load_vector}).
Since $\mathcal A$ is a sparse matrix, the system can be efficiently solved by an iterative method (such as GMRES) and an appropriate pre-conditioner 
in many cases \citep{Hack.LAS}.

However, as mentioned by \citet{Kanschat2004Preprint} the matrix $\mathcal A$ is sometimes badly conditioned, especially in cases with high opacity 
and single 
scattering albedos. The usual pre-conditioners (such as ILU \citep{Hack.LAS}) in combination with GMRES then fail to yield a converged solution. In 
this case a more adapted pre-conditioner is required to solve the system of equations \eqref{eq:problem_lse}. An example of such a pre-conditioner 
based on the Eddington approximation but limited to a plane-parallel situation is given by \citet{Kanschat2004Preprint}. Nonetheless such a pre-conditioner should be adaptable 
also to spherically symmetric atmospheres which will be subject of a forthcoming publication.

\section{Test Problems}
\label{sec:test_problems}

In this section, we test our previously described numerical solution of the radiative transfer equation with a discontinuous Galerkin method. 
Caused by the availability of exact solutions, we need to rely on test problems for comparison, where the resulting radiation field is known a 
priori, or compare with results of previously published calculations.

The first two tests evaluate the numerical stability and flux conservation in an empty atmosphere. In this case, the photons follow the 
characteristic paths of the transfer equation \eqref{eq:rte_1}. These scenarios also feature discontinuities in their solutions, which will serve as a 
test for the numerical stability of our DG formulation.

For the third scenario, we refer to the radiative transfer calculations for an isotropically scattering, spherical atmosphere with a power-law 
extinction coefficient \citep{Hummer1971MNRAS}. We compare our results with 
their published values and also verify that the energy is also conserved in this case as required.
Note, that the intention of these calculations is to validate  and verify the numerical approach and its capabilities to cope with e.g. 
discontinuities or highly scattering atmospheres.

\subsection{Computational details}

We implemented the numerical algorithms described in the previous section by using the general finite element library GetFEM++\footnote{http://home.gna.org/getfem/}.
In principle, the element-wise calculations can also be performed in sense of \citet{Alberty1999NuAlg..20..117A}.
Details on our numerical implementation are given in the Appendix.

A DG(1) method on a structured grid with rectangular $Q_1$ elements is used for the first two test problems. For both cases, the grid 
contains in total 100 grid points in the radial direction as well as 80 grid points in the angular direction, respectively. 
The radius grid points are distributed equidistantly, while the angular mesh is constructed by using a Gaussian quadrature rule. 
The relatively high resolution is required to capture the discontinuous solutions properly. 
In principle, a much lower resolution could be used in these cases, if an unstructured grid is 
constructed, where the elements are directly aligned to capture the (previously) known discontinuities in the solution.

For the third test problem, we use a DG(2) method on a structured $Q_2$ grid with 25 radial points and 10 angular values. The radial nodes 
are placed in logarithmic equidistant steps, while for the angular grid a Gaussian quadrature rule is again employed.

The resulting linear system of equations is solved by an iterative GMRES solver with a restart value of 10 in each case. The iteration matrix 
is preconditioned by an incomplete LU factorisation.

\subsection{Empty atmospheres with an isotropically emitting inner surface}

\begin{figure}
  \centering
  \includegraphics[scale=0.70]{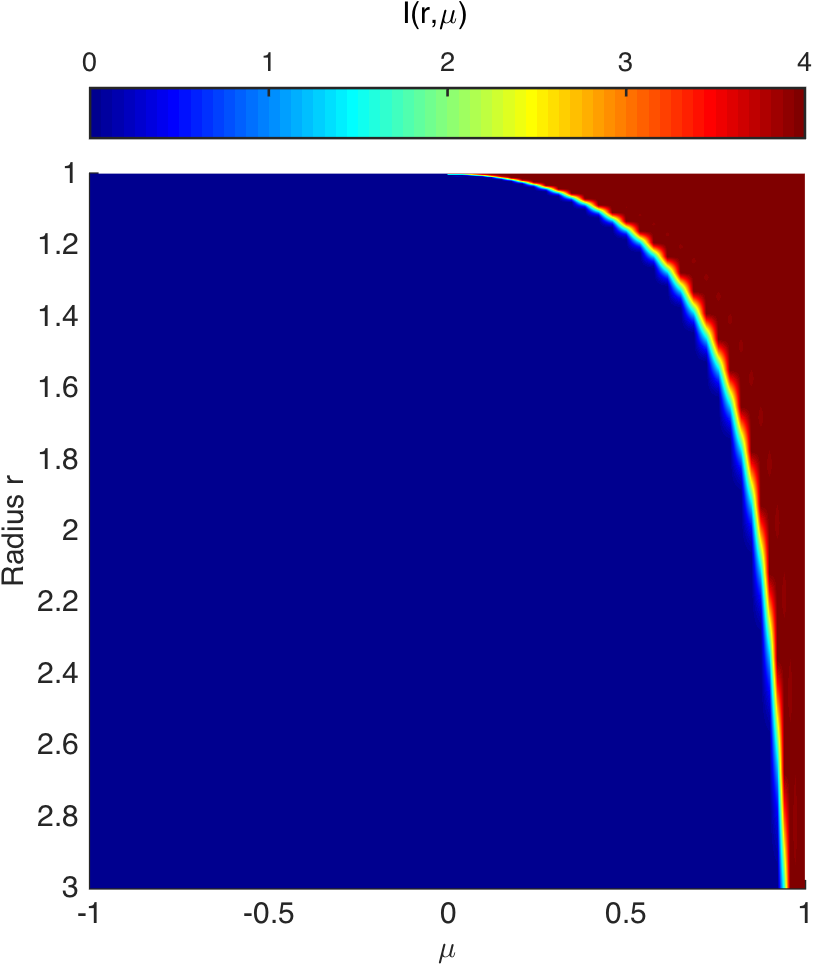}
  \caption{Radiation field $I(\mu,r)$ as a function of radius $r$ and angular variable $\mu$ for an empty, extended atmosphere with an isotropically 
emitting inner surface.}
  \label{fig:result_problem1}
\end{figure}

As a very first test we study the following simple radiative transfer problem:
\begin{equation}
  \begin{aligned}
	&r_\mathrm{in} = 1.0,\\
	&r_\mathrm{out} = 3.0,\\
	&\kappa(r) = \eta^{\text{sp}}(r) = \eta^{\text{ind}}(r) = \overline{s}(r) = 0
	\end{aligned}
\end{equation}
with boundary conditions
\begin{equation}
  \begin{aligned}
	  &I_{\text{in}} = 4,\\
	  &I_{\text{out}} = 0,
	\end{aligned}
\end{equation}
which represents an empty, extended atmosphere with an isotropically emitting inner surface.

\begin{figure}
  \centering
  \includegraphics[scale=0.60]{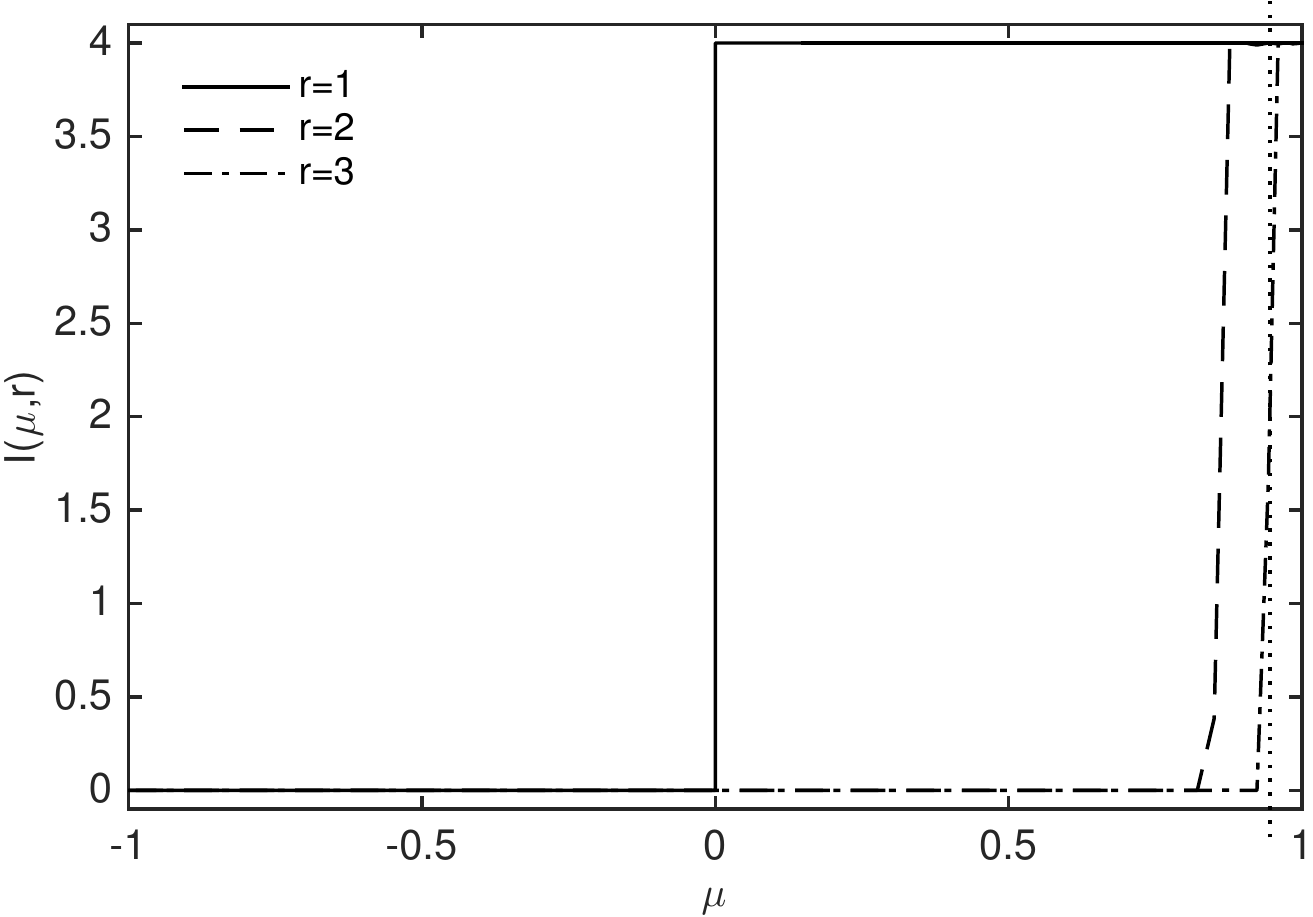}
  \caption{Radiation field $I(\mu,r)$ as a function of the angular variable $\mu$ and three chosen radii $r$. The vertical dotted line denotes 
$\mu_c$, the 
limiting value of $\mu$ under which the inner emitting surface is seen at the outer boundary $r = 3$. Note, that the discontinuity is accurately 
captured (i.e. the solution shows no oscillations).}
  \label{fig:result2_problem1}
\end{figure}

The resulting radiation field $I(\mu,r)$ is shown in Fig. \ref{fig:result_problem1}. Additionally, several cuts through the parameter space are given 
in Fig. \ref{fig:result2_problem1}, where the intensity $I$ is shown as a function of $\mu$ for three different values of $r$. Note, that in this 
radiative transfer problem a real discontinuity is present in the solution which separates the region of constant intensity originating from the inner 
surface and the regions where the intensity is zero. The characteristic separating these two regions is given by 
\begin{equation}
  \mu_\text{c} = \sqrt{1-\frac{r_\text{in}^2}{r^2}} \ .
	\label{eq:characteristic_inner_boundary}
\end{equation}
The results in Figs. \ref{fig:result_problem1} and \ref{fig:result2_problem1} clearly show the spherical dilution of the outward directed intensity. 
With increasing radius $r$ the angle $\mu_\text{c}$, under which the central emitting surface can be seen, decreases strongly. This is illustrated in 
greater detail in Fig. \ref{fig:result2_problem1}, where the intensity $I$ is shown as a function of $\mu$ and for three different values of $r$. The 
vertical dotted line marks the limiting value of $\mu_\text{c}(r_\text{out})$ under which the emitting inner surface is seen at the outer boundary. 
As 
suggested by Fig. \ref{fig:result2_problem1} the DG method presented here is accurately reproducing this spherical dilution of the radiation field 
without introducing numerical oscillations.

Consequently, the DG method is capable of capturing this discontinuity quite well without introducing numerical oscillations which would occur in 
case of a 
continuous Galerkin method. In practice, the discontinuity is smeared out over one element. In order to limit the errors introduced by this 
out-smearing one could either use a high mesh resolution in general, or use an adaptive mesh refinement along the discontinuous boundaries of the 
intensity given by Eq. \eqref{eq:characteristic_inner_boundary}.

Because there are no interactions of the atmosphere with the radiation field in this test example, the radiative transfer problem should be 
automatically energy conserving. For a spherically symmetric atmosphere this means that the flux $(r^2 H(r))$ is constant throughout the atmosphere, 
where $H$ is the first angular moment (Eddington flux) of the intensity
\begin{equation}
  H(r) = \frac{1}{2} \int_{-1}^{+1} I(r,\mu) \, \mu \, d \mu \, .
\end{equation}
In this first test calculation, the flux is conserved better than $0.01\%$ everywhere in the computational domain, indicating that the DG method is 
stable and accurate for such a problem.

\subsection{Empty atmospheres, illuminated by an exterior light source}

This test case differs from the previous one only by choosing different boundary conditions
\begin{equation}
  \begin{aligned}
	  &I_{\text{in}} = 0 ,\\
	  &I_{\text{out}} = 4 \mu \, .
	\end{aligned}
\end{equation}

\begin{figure}
  \centering
  \includegraphics[scale=0.70]{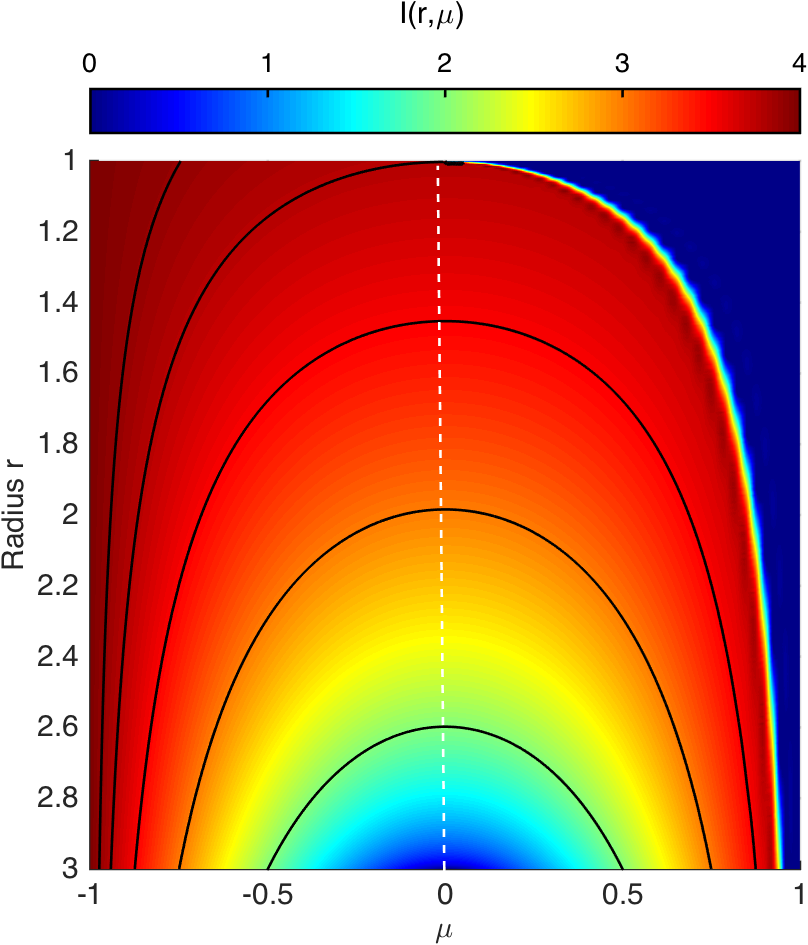}
  \caption{Radiation field $I(\mu,r)$ as a function of radius $r$ and angular variable $\mu$ for an empty, extended atmosphere, illuminated by an 
exterior light source. The curved black lines illustrate a number of selected characteristics of this radiative transfer problem, i.e. the pathways of 
photons through the atmosphere. The vertical white line separates in and outgoing directions.}
  \label{fig:result1_problem2}
\end{figure}

\begin{figure}
  \centering
  \includegraphics[scale=0.60]{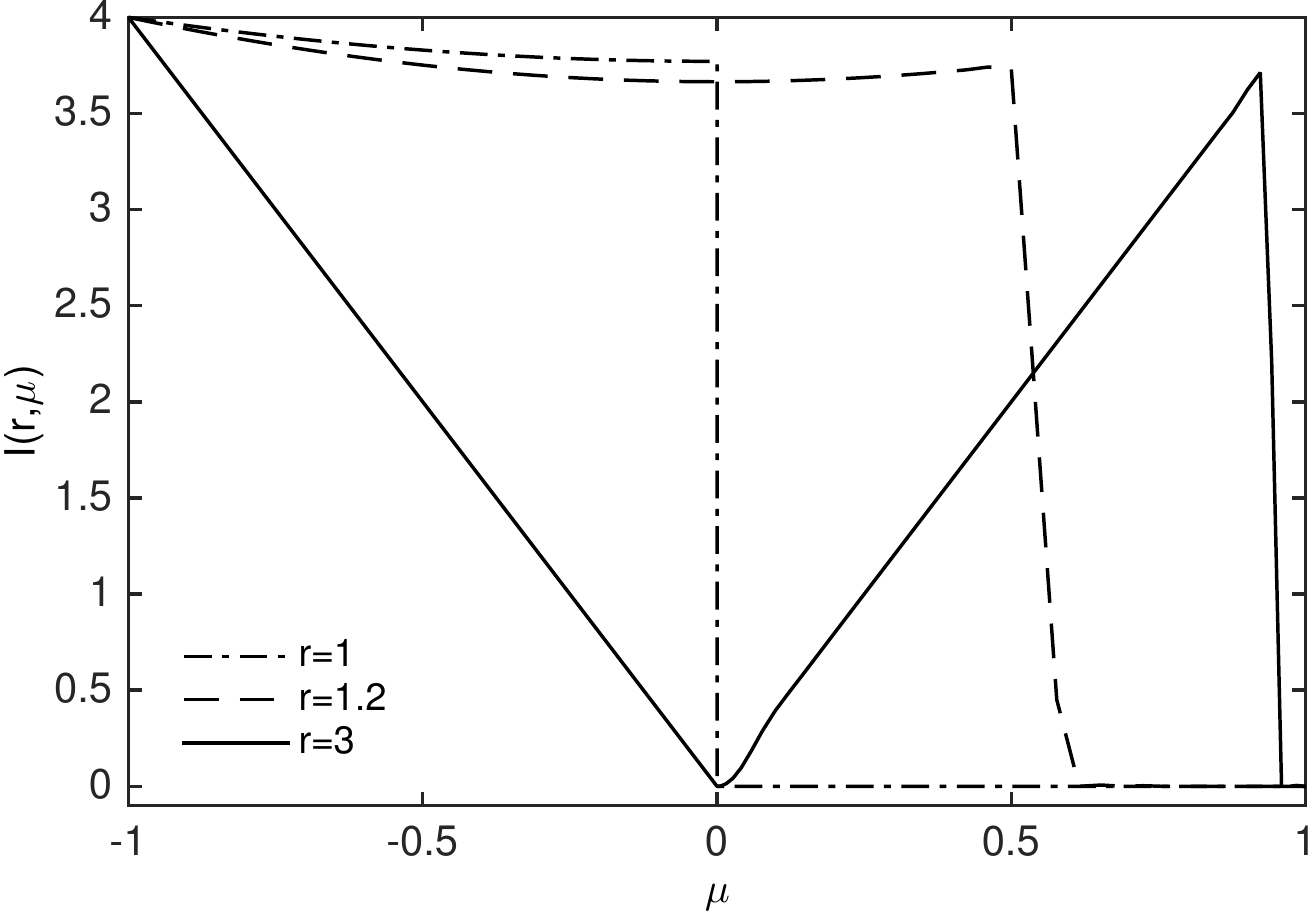}
  \caption{Radiation field $I(\mu,r)$ as a function of the angular variable $\mu$ and three chosen radii $r$. Note, that the discontinuity is 
accurately captured (i.e. no oscillations in the solution).}
  \label{fig:result2_problem2}
\end{figure}

This problem then represents an empty, extended atmosphere illuminated by an exterior light source. The particular form of the outer boundary 
condition has 
been chosen to highlighted the photons' characteristics, i.e. the geometric pathways of photons travelling through the atmosphere. The resulting 
radiation field is shown in Figs. \ref{fig:result1_problem2} and \ref{fig:result2_problem2}. 

As it can clearly be noticed from Fig. \ref{fig:result1_problem2}, only a small fraction of the incident radiation eventually reaches the inner 
surface. The remaining radiation is transmitted through the atmosphere and leaves the computational domain at the outer boundary with the same 
intensity and under the same angle as the incident radiation field. The curved black lines in Fig. \ref{fig:result1_problem2} illustrate some of the 
characteristics of this radiative transfer problem, i.e. the pathways of photons through the atmosphere. This behaviour is very different in 
comparison to problems with 
plane-parallel geometry, where the incident light under any angle would eventually reach the surface.

Like the previous test case, this radiative transfer problem also has a discontinuous solution in a part of its computational domain. This 
discontinuity is also captured accurately by the DG formulation introduced in this work (see Fig. \ref{fig:result2_problem2}). Again, the flux is conserved better than $0.01\%$ everywhere in the computational domain.

\subsection{Isotropically scattering atmosphere}

To prove the capability of our DG radiative transfer method to handle non-local scattering problems, we compare in this subsection our results with 
calculations published in \citet{Hummer1971MNRAS} resembling a purely scattering, spherical atmosphere with a power-law opacity. In contrast to our 
work, \citet{Hummer1971MNRAS} used an iterative moment method to obtain the solution of the radiative transfer equation. Due to the limitation of this 
moment method, only isotropic scattering was taken into account by \citet{Hummer1971MNRAS}. The radiative transfer problem considered here for 
comparison is therefore specified by
\begin{equation}
  \begin{aligned}
  &r_\mathrm{in} = 0.01,\\
  &r_\mathrm{out} = 0.1,\\
  &\kappa(r) = \eta^{\text{sp}}(r) = \eta^{\text{ind}}(r) = 0,\\
  &\overline{s}(r) = r^{-3/2} \, .
  \end{aligned}
\end{equation}
Following \citet{Hummer1971MNRAS}, scattering is assumed to be isotropic, i.e. the scattering phase function is given by
\begin{equation}
  p^{(0)}(\mu,\mu') = 1 \, .
\end{equation}
Boundary conditions are chosen to resemble those introduced in \citet{Hummer1971MNRAS}. The moment method employed by \citet{Hummer1971MNRAS} allowed 
for a direct prescription of the radiation flux at the inner boundary, which was set to $r_\text{in}^{2} H(r_\text{in}) = 1$. At the outer boundary, 
no incident radiation was assumed. To replicate these constant flux boundary conditions, we employed the following conditions on the specific 
intensity at the domain boundaries
\begin{equation}
  \begin{aligned}
    &I_{\text{in}} = 4 \, r_\text{in}^{-2} - 2 \int_{-1}^{0} \mu \, I(\mu,r_\text{in}) \, d\mu\\
    &I_{\text{out}} = 0
  \end{aligned}
\end{equation}
This inner boundary condition is obtained from the requirement that
\begin{equation}
  \begin{split}
  r_\text{in}^{2} H(r_\text{in}) &= r_\text{in}^{2} \frac{1}{2} \int_{-1}^{+1} \mu \, I(\mu,r_\text{in}) \, d\mu\\ 
	                                   &= r_\text{in}^{2} \frac{1}{2} \int_{-1}^{0} \mu \, I(\mu,r_\text{in}) \, d\mu 
					    + r_\text{in}^{2} \frac{1}{2} \int_{0}^{+1} \mu \, I(\mu,r_\text{in}) \, d\mu\\
				           &= r_\text{in}^{2} \frac{1}{2} \int_{-1}^{0} \mu \, I(\mu,r_\text{in}) \, d\mu 
					    + r_\text{in}^{2} \, \frac{1}{4} \, I(+\mu,r_\text{in}) \\
					   & = 1\, ,
  \end{split}
\end{equation}
assuming an isotropic distribution of $I(+\mu,r_\text{in})$.

\begin{figure}
  \centering
  \includegraphics[scale=0.70]{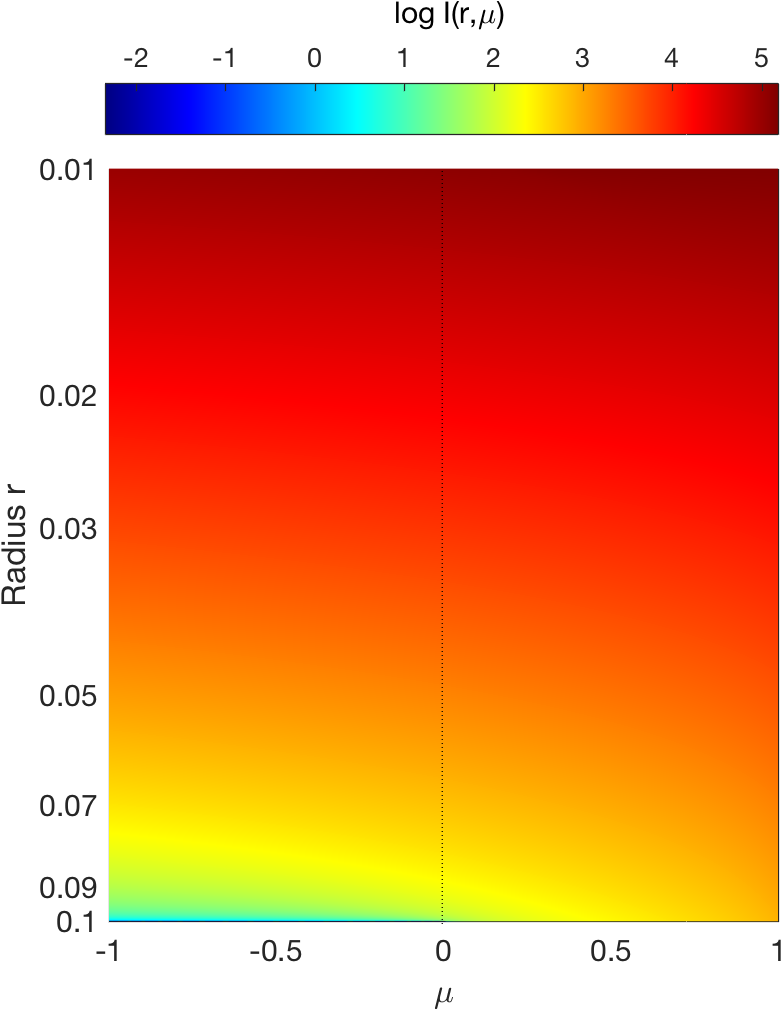}
  \caption{Radiation field $I(\mu,r)$ as a function of radius $r$ and angular variable $\mu$. The vertical black line separates in and outgoing 
directions.}
  \label{fig:result1_problem3}
\end{figure}

Because of the purely scattering atmosphere, the radiation flux $r^2H(r)$ needs to be conserved throughout the atmosphere. The resulting values of 
$r^2 H(r)$ differ from the assumed constant flux at the inner boundary $r_\text{in}^2 H(r_\text{in}) = 1$ by less than 0.5\% everywhere in the 
computational domain. In the inner part of the atmosphere, the highest deviations are below $0.001\%$ indicating again the capability of the DG 
method.

The resulting radiation field $I(\mu,r)$ is shown in Fig. \ref{fig:result1_problem3}. In contrast to the previous two test problems, no discontinuity 
occurs in this case. Due to the many multiple scattering events, the radiation field exhibits a smooth behaviour in angle and radius. At the inner 
boundary, the radiation field is isotropic, resulting in an Eddington factor of 0.3 as expected.

\paragraph{Mean intensity}

\begin{figure}
  \centering
  \includegraphics[scale=0.60]{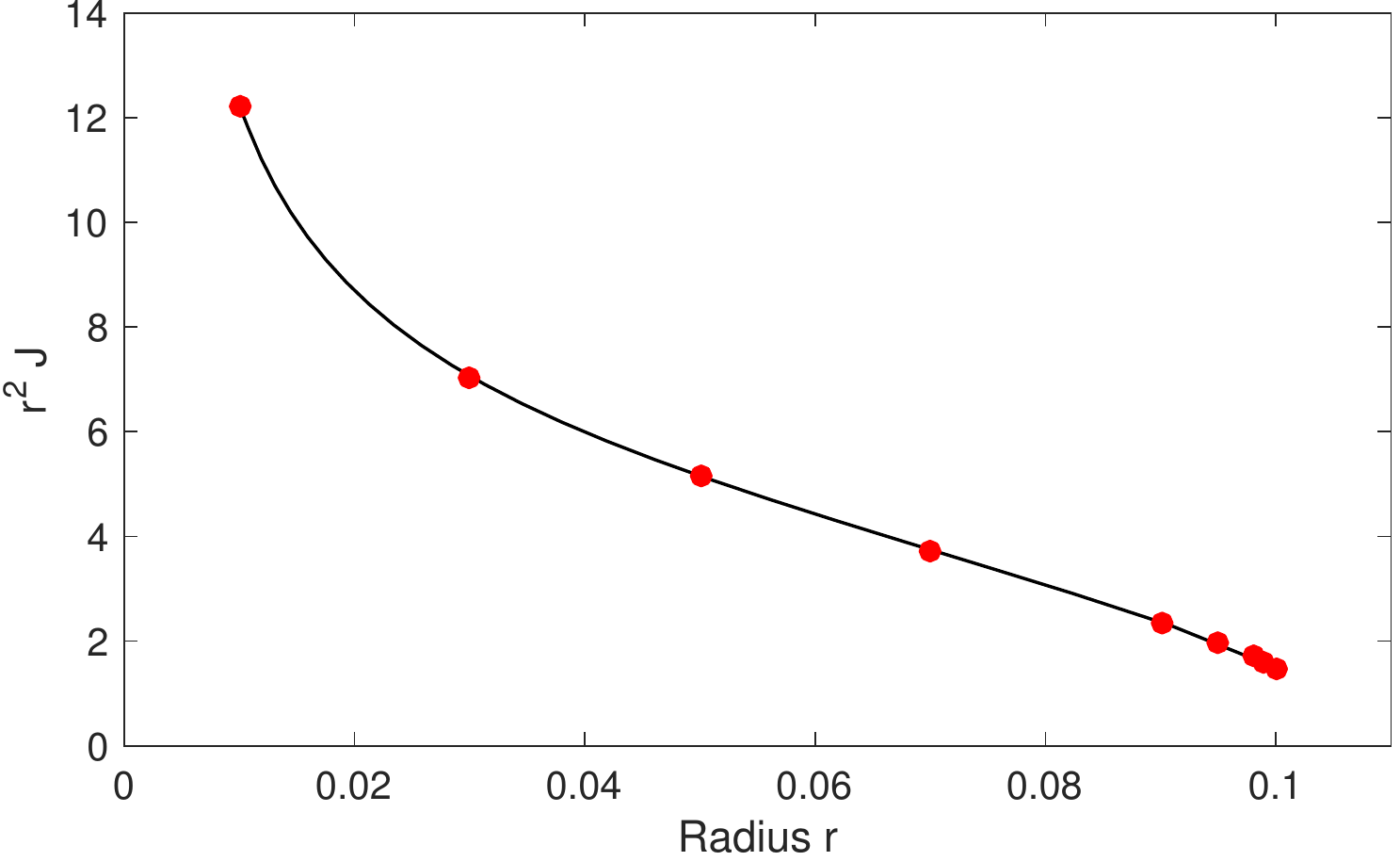}
  \caption{Mean intensity $r^2 J$ as a function of the radius for an isotropically scattering atmosphere. The dots mark the corresponding results 
published by \citet{Hummer1971MNRAS}.}
  \label{fig:res3}
\end{figure}

Figure \ref{fig:res3} shows the resulting values of the mean intensity $r^2 J(r)$ in comparison to the corresponding tabulated values taken from 
Table II in \citet{Hummer1971MNRAS}. Agreement with the results of \citet{Hummer1971MNRAS} are excellent in all parts of the isotropically 
scattering atmosphere. It should be noted, that the grid resolution used in the present calculation is much lower than the one employed by 
\citet{Hummer1971MNRAS} (200 $r$-points, 100 angles). We use a factor of eight less radius grid points and a factor of ten less angular values. The 
radiation field as a function of radius and angle is rather smooth and, thus, does not require a high resolution -- especially since the intensity 
here is assumed to be a second-order polynomial within each element.

\paragraph{Limb-darkening}

\begin{figure}
  \centering
  \includegraphics[scale=0.60]{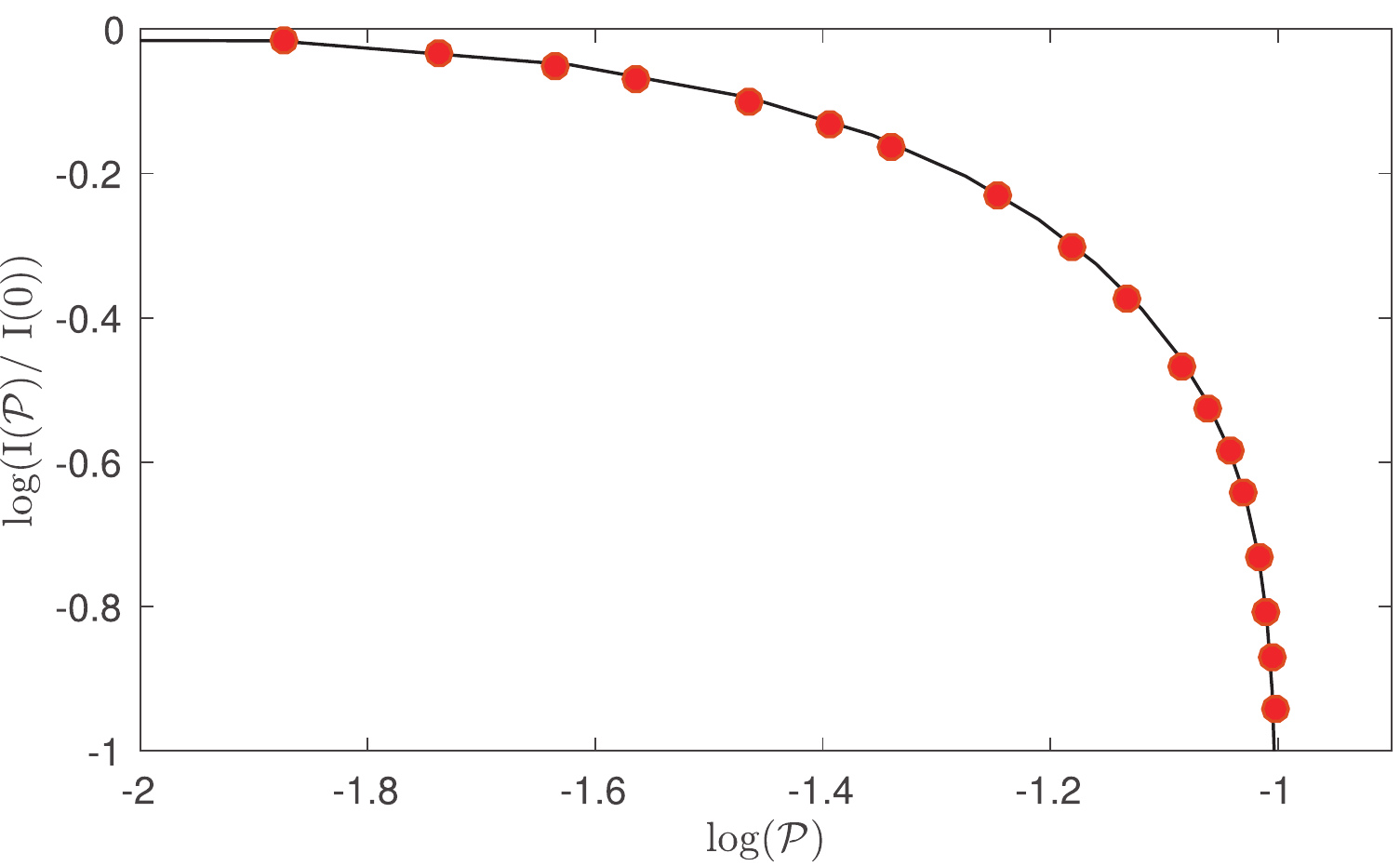}
  \caption{Limb-darkening curve $I(p)/I(0)$ as a function of the impact parameter $\mathcal P$. The dots mark the corresponding results from 
\citet{Hummer1971MNRAS}, which have been directly read off their Fig. 6.}
  \label{fig:limb_darkening}
\end{figure}

The limb-darkening curves $I(\mathcal P)/I(0)$ are shown in Fig. \ref{fig:limb_darkening} as a function of the impact parameter $\mathcal P$, given by

\begin{equation}
  \mathcal P = r_{\text{out}} \sqrt{1 - \mu^2} \, .
\end{equation}

The curve shows an excellent agreement with the results of \citet{Hummer1971MNRAS} (their Fig. 6). The calculated disk centre intensity $I(0)$ is 
$816.4$, which deviates by less than $0.5\%$ from the value of $820$ given in \citet{Hummer1971MNRAS}. Note, however, that their solution was stated 
to have a flux conservation error of a few percent at the outer boundary, while our flux is conserved better than $0.5\%$ everywhere, which can 
explain the slight deviations in the disk centre intensity.

\section{Numerical efficiency}
\label{sec:efficiency}

In this section, we study the development of the computational time as a function of the total number of degrees of freedom $N$ by using the 
previous, third test example for the DG(2) method (isotropic scattering atmosphere). 
Thereby, we test both, the dependence on the number of angular and radius points, separately. 
Successively increasing the resolution in either grid direction, we track the development of the computational effort. For a grid composed of 
$Q_q$ elements with $N_q$ local degrees of freedom $N$ is given by
\begin{equation}
	N = (N_r - 1) \cdot (N_\mu - 1) \cdot N_q \ ,
\end{equation}
where $N_r$ and $N_\mu$ are the number of radial and angular grid points. 
According to
\begin{equation}
N_q = (q+1)^2 \ ,
\end{equation}
the number of local degrees of freedom for the $Q_2$ elements used here is nine (see Fig. \ref{fig:P_q_Q_q}).
Note that the computational time includes every single step to solve the problem, from the construction of the grid, finding the corresponding elements 
throughout the grid that need to be connected by the scattering integral, calculating the element-wise contributions to the matrix $\mathcal A$, and 
solving the linear system of equations. In reality, many of these tasks need to be done only once -- such as the element connections -- and their 
results can be re-used at different iteration steps or for other frequencies.

\begin{figure}
  \centering
  \includegraphics[scale=0.58]{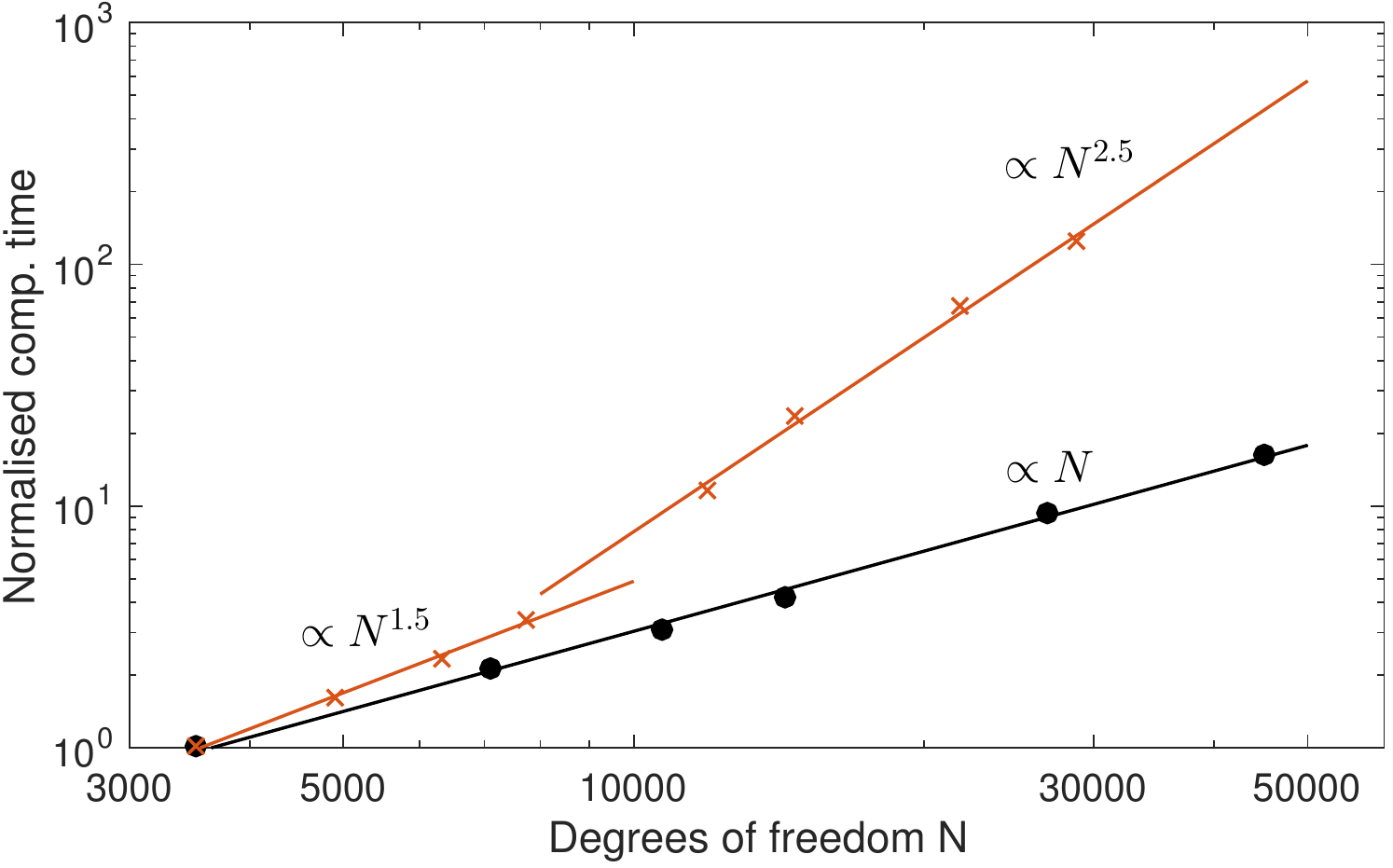}
  \caption{Development of the normalised computational time as a function of the number of degrees of freedom. Both variables are depicted in
logarithmic scale.
  The black dots mark the results for increasing the number of radius grid points from 40 to 500, while
  the number of angles is kept constant at 10. 
  The red crosses indicate the results for increasing the angular resolution from 10 to 82, using a constant radial grid point number of 40.
  The corresponding solid lines depict the linear fits through the respective data points.
  For the angular nodes, two different fits are used.}
  \label{fig:computing_time}
\end{figure}

The corresponding results are shown in Fig. \ref{fig:computing_time}, which also depicts the corresponding linear fits through the obtained data 
points to determine the order of dependence on the grid point number. The results in Fig. \ref{fig:computing_time} suggest that the computing time 
increases linearly with the number of radius points. For smaller number of angles, the computational time increases with $N^{1.5}$. For numbers 
of $N$ larger than about 10000, this dependency changes to $N^{2.5}$ (see Fig. \ref{fig:computing_time}). 
For comparison, the iterative method using variable 
Eddington factors as employed by \citet{Hummer1971MNRAS} typically scales cubic in the number of radius grid points for one iteration step 
\citet{Mihalas1978}, if for each radius point a corresponding tangent ray (equal to two angular directions) is used. Several iterations, however, are 
needed to obtain a converged solution, which means that the overall computational cost is  
larger than this estimate.

It should be noted, that with increasing numbers of degrees of freedom, most of the time is spent solving the linear system of equations. 
This is, in fact, the main reason for the increase in the computational time for the angular points. The scattering integral 
(Eq. \eqref{eq:scattering_integral}) is responsible for connecting elements in the final matrix $\mathcal A$ which are, however, not direct 
neighbours in the grid. 
Depending on the actual implementation of the element numbering scheme (and, thus their position in $\mathcal A$), the contributions to the integral 
Eq. \eqref{eq:scattering_integral} by those elements can be scattered throughout the matrix $\mathcal A$. 
This makes an efficient solution of the linear system of equations quite hard for an iterative solver. 
A considerable decrease in computing time could, therefore, be obtained when the element numbering scheme would take the elements' connections due to 
the scattering integral into account, such that the contributions of these elements would be located closer to the main diagonal of the matrix 
$\mathcal A$ \citep[see e.g.][]{Cuthill:1969:RBS:800195.805928}. For a purely scattering atmosphere, we were able to obtain converged 
solutions up to optical depths of around 100. Beyond that, the matrix $\mathcal A$ becomes too badly conditioned to be solved by a standard GMRES 
scheme with an ILU preconditioner. As mentioned in Sect. \ref{sec:transformation_matrix}, a more sophisticated preconditioner needs to be 
developed for such cases.

\section{Outlook}
\label{sec:outlook}

The implementation of the DG-FEM presented in this work can be greatly extended in the future in view of its numerical efficiency.
One improvement which can be addressed is adaptive grid refinement in combination with error estimation. 
To do so, the residual in Eq. \eqref{eq:element_res2} provides the opportunity to obtain an a posteriori error estimate. 
The point wise error of the calculated solution can be estimated by expressing this equation in a suitable norm.
In regions, where the error is above a certain threshold, a local 
grid refinement or a local increase of the polynomial order can then be used to increase the numerical accuracy. 
Using this adaptive grid refinement ensures, that the exact solution is calculated accurately on the entire grid. 
Adaptive mesh refinement can become important when the solution features a 
non-monotonous intensity distribution. This can occur within circumstellar dust shells, for example, where the dust forming region somewhere in the 
atmosphere creates a local maximum in the angular intensity distribution \citep{Gail2014pccd.book.....G}.  
Such a non-monotonous behaviour can easily be traced by local grid refinement, without increasing the resolution in other parts of the grid.

Another option to improve the numerical efficiency of the presented DG-FEM is to parallelise the numerical work. In principle, 
since the DG-FEM formulation does not enforce continuity across element boundaries, all element-wise calculations presented in Sect. 
\ref{sec:elment_calculations} can be computed independently from each other. 
Thus, such independent computations can be parallelised very easily by using MPI or OpenMP schemes, for example. 
Furthermore, the calculations within each element -- unless pushed to very high orders -- are always a comparatively small 
numerical problem. 
Such small problems are ideal to be run on GPU clusters, i.e. graphic card processors with thousands of cores, which could increase 
the computational speed by at least one order of magnitude.

Finally, the solution of the final linear system of equations can also be parallised. Special libraries, such as PETSc, are available to 
distribute the computational task across several CPUs or GPUs. This would be especially helpfull in cases, where most of the time is spent by the 
iterative solver.

\section{Summary}
\label{sec:summary}

In this work we presented a numerical method to directly solve the radiative transfer equation in spherical symmetry. We used a discontinuous Galerkin 
finite element method to discretise the transport equation as a two-dimensional boundary value problem. Arbitrary complex scattering phase function 
can, in particular, be integrated exactly using this approach. The discretisation yields a sparse system of equations which can be solved by standard 
iterative methods with suitable preconditioners. 

The applicability of the DG-FEM is verified on a set of different test problems. The discontinuous Galerkin method is able to directly describe the 
spherical dilution of the radiation field in an extended atmosphere. We show that the DG-FEM is able to accurately capture discontinuities which 
might appear in the solution of the transfer equation for special test problems. We also compared our method with previously published radiative 
transfer calculations for a spherical, isotropically scattering atmosphere. The agreement between the different approaches was found to be excellent. 
Consequently, the discontinuous Galerkin method is perfectly suitable to treat accurately the radiative transfer in a spherical atmosphere even with e.g. 
discontinuities in the radiation field or complex scattering behaviour, which might cause severe numerical problems for other radiative transfer 
solution methods.

\begin{acknowledgements}
      D.K gratefully acknowledges the support of the Center for Space and Habitability of the University of Bern and the MERAC Foundation for partial financial assistance.
      This work has been carried out within the frame of
      the National Centre for Competence in Research PlanetS supported by the Swiss
      National Science Foundation. D.K. acknowledges the financial support of
      the SNSF. The authors thank the anonymous referee for the very constructive and helpful comments
      that greatly improved the manuscript.
\end{acknowledgements}

\bibliographystyle{aa}
\bibliography{dg}

\begin{appendix}

\onecolumn

\section{Implementation details}

In this Appendix, we briefly give details on our implementation of the DG-FEM in a numerical code as one example. As previously mentioned, we use the 
general finite element library 
GetFEM++. GetFEM++ provides a large set of high-level functions and tools which can assist in easily setting up a finite element method. A complete 
description of the library can be found in its user documentation\footnote{http://home.gna.org/getfem/}. In this appendix we only include the basic 
set of necessary steps. We refer the reader to GetFem++ documentation for much more detailed descriptions and explanations on the contents and 
implementations of the library.

If, alternatively, one is interested in building a finite element from scratch, we refer to e.g. \citet{Hesthaven2008nodal}, 
\citet{Eriksson1996cde..book.....E}, or \citet{Alberty1999NuAlg..20..117A} for more details.

\subsection{Grid and element generation}

The first step in setting up a finite element method is the generation of the grid and the definition of the elements.
Within GetFEM++, the mesh generation can be done via different approaches. One possibility is to construct a grid outside of GetFEM++ (for example by 
using the mesh generator Gmsh) and import it. It is also possible to use GetFEM++ itself for grid generation. The built-in mesh generation tools, 
however, cover only very simple grids. A third possibility is to construct a special grid by defining all vertices of the mesh's elements. We use 
this last option since we, for example, want to distribute the angular vertices according to a Gaussian quadrature rule which is not covered by 
GetFEM++'s provided mesh tools.

A single point is added to a mesh cell with\\

\noindent \texttt{pts[i] = bgeot::base\_node(radius\_point,angular\_point);} ,\\

\noindent followed by\\

\noindent \texttt{mesh.add\_parallelepiped\_by\_points(2, pts.begin());}\\

\noindent to create, e.g., a square mesh cell.

For a given mesh, GetFEM++ will then automatically create the elements. It offers a wide choice of different element definitions (see the 
user documentation for a complete overview). 
For our DG-FEM code, we use Lagrangian elements (see Fig. \ref{fig:P_q_Q_q}), which are created via\\

\noindent \texttt{mesh\_fem.set\_finite\_element(getfem::fem\_descriptor("FEM\_QK\_DISCONTINUOUS(2,2)"));} ,\\

\noindent for a second-order method on a two-dimensional grid. This has to be coupled with a suitable quadrature method, e.g.\\

\noindent \texttt{mesh\_integration.set\_integration\_method(getfem::int\_method\_descriptor("IM\_GAUSS\_PARALLELEPIPED(2,4)"));}\\

\noindent GetFem++ also offers other quadrature methods which are extensively described in the library's documentation.

Given a mesh and the definition of the corresponding elements of a chosen polynomial degree, GetFEM++ automatically produces the transformation 
matrix $\mathcal T_{\alpha i}^{k}$ to map the local coordinates $i$ within the $k$-th element to global coordinates $\alpha$. 
GetFEM++ also provides tools to perform calculations of base functions and their derivatives, as well as doing only computations on the faces of 
elements and providing the corresponding normal vectors at the elements' faces.

\subsection{Mass matrix}

Within GetFEM++, all computations are usually defined on a reference element. The library provides high-level methods for the assembly of the mass 
and stiffness matrices. For example, the computation of the mass matrix on a single element (cf. Eq. \eqref{eq:mass_matrix})
\begin{equation}
  \mathcal M_{ij}^k := \int_{\D^k} \left( \frac{2\, \mu}{r} + \hat{\chi} \right)\,  \ell^k_j \, \ell^k_i d\x
  \label{eq:appendix_mass_matrix}
\end{equation}
is simply defined by\\

\noindent \texttt{"a=data(\#1);" "M(\#1,\#1) += a(i).comp(Base(\#1).Base(\#1).Base(\#1))(i,:,:)"} ,\\

\noindent where \texttt{Base(\#1)} are the base functions $\ell^k$ and \texttt{a} contains the function
\begin{equation}
  a = \frac{2\, \mu}{r} + \hat{\chi} \ ,
\end{equation}
which, following Eq. \eqref{eq:sol_exp_local}, is expanded into
\begin{equation}
a(\x) = \sum_{i=1}^{N_p} a(\x_i^k) \,\ell_i^k(\x) .
\end{equation}
This summation is symbolised with the \texttt{i} in the definition \texttt{(i,:,:)}. The remaining two, free indices (denoted by \texttt{:}) are the 
corresponding matrix indices.

The global mass matrix is defined according to\\

\noindent \texttt{mass\_matrix.set("a=data(\#1);" "M(\#1,\#1) += a(i).comp(Base(\#1).Base(\#1).Base(\#1))(i,:,:)");}\\
\noindent \texttt{mass\_matrix.assembly();} ,\\

\noindent where the local contributions of each element are calculated and then automatically mapped to the real elements by means of 
a Jacobian matrix.

\subsection{Stiffness matrix}

In analogy to that, the local stiffness matrix (cf. Eq. \eqref{eq:stiffness_matrix_r})
\begin{equation}
  \mathcal S^k_{ij,r} := - \int_{\D^k} \mu\, \ell^k_j \,\frac{\partial \ell^k_i}{\partial r} \, d\x
  \label{eq:appendix_stiffness_matrix}
\end{equation}
is defined by\\

\noindent \texttt{"M(\#1,\#1) -= comp(Base(\#2).Base(\#1).Grad(\#1))(1,:,:,1);"}.\\

\noindent The parameter \texttt{Base(\#2)} describes the $\mu$ factor in Eq. \eqref{eq:appendix_stiffness_matrix}, which in the context of GetFem++ 
is expressed by a global function. This function returns the $\mu$ value for any given point $\x$. The derivative of the base function with respect 
to $r$ is denoted by \texttt{Grad(\#1)}. The \texttt{1} in the last entry in the definition \texttt{(1,:,:,1)} signifies that the derivative has to 
be performed with respect to the first dimension of the two-dimensional grid - which, in our case, is the radius coordinate.
The global stiffness matrix is then assembled with\\

\noindent \texttt{stiffness\_matrix.set("M(\#1,\#1) -= comp(Base(\#2).Base(\#1).Grad(\#1))(1,:,:,1);");}\\
\noindent \texttt{stiffness\_matrix.assembly();} .\\

Of course, the second stiffness matrix (Eq. \eqref{eq:stiffness_matrix_mu}) is calculated analogously.

\subsection{Numerical flux}

The numerical flux (Eq. \eqref{eq:dg_jump}) is evaluated by integrating along the faces of each element, taking into account the contributions of adjacent elements:
\begin{equation}
    \mathcal F^{kn}_{ij} := \int_{\partial \D^k} \hat{\boldsymbol n}_r \cdot (f)^*\!\left(I_h^k(\ell_i^k),I_h^n(\ell_j^n)\right) \, \ell_i^k d\x \ .
\end{equation}
Since the high-level methods from the previous subsection only operate on one local element, one has to use the special
\texttt{getfem::compute\_on\_inter\_element} class which also includes information on the properties of elements sharing a common face. The class' 
method \texttt{compute\_on\_gauss\_point} has to be carefully adapted to calculate the contribution to the integral of Eq. \eqref{eq:dg_jump} at each 
of the 
integral's quadrature nodes. Within this special object class, GetFEM++ provides all necessary information to evaluate the integral, such as the 
normal vectors, quadrature nodes and weights along the face, as well as the entries of the Jacobian matrix to map the computations on the reference 
element 
back to the real one.
More details on the implementation of this class can be found in the official documentation and the corresponding code examples published together 
with the source code of the library.

\subsection{Scattering matrix}
The scattering matrix is given by (cf. Eq. \eqref{eq:scattering_integral})
\begin{equation}
  \mathcal J^{kn}_{ij} = \int_{\D^k} \left( \frac{\overline{s}}{2} \int_{-1}^{+1} \ell_j^n(r,\mu') \, p^{(0)}(r,\mu,\mu')\,  d\mu' \right) \, \ell_i^k(\x)\,  d\x \ .
\end{equation}
It contains two integrals, the outer one for the integration over a local element and the inner one which integrates the scattering phase function 
over multiple different elements. For this more complicated term GetFEM++ also provides no direct high-level method.

We use a Gaussian quadrature rule for the discretisation of the inner integral, with an order high enough to ensure an exact integration 
of the phase function. At each of the Gaussian quadrature nodes from this scattering integral $\mu'$, one needs the value of the base function 
$\ell_j^n(r,\mu')$, which usually lies outside of the considered, local element $k$. Unless a degree of freedom is directly located at the 
coordinates $(r, 
\mu')$, the base function $\ell_j^n(r,\mu')$ contains contributions from several different degrees of freedom within the element $n$. 

GetFEM++ provides interpolation routines to find the corresponding values of $\ell_j^n$ for a given radius $r$ and angle $\mu'$ as well as the set of 
degrees of freedom which contribute to this specific base function.
Given the mapping from the local coordinates of $i$-th degree of freedom in the $k$-th element to the global coordinate $\alpha$ and the 
corresponding mappings of all degrees of freedom connected by the scattering integral with global indices $\beta$, one can directly add their 
contributions to the global matrix $\mathcal A_{\alpha \beta}$.

\subsection{Boundary conditions}

The boundary conditions from Eq. \eqref{eq:dg_method} can be implemented in two different ways. They can either be introduced in analogy to the 
integral of the numerical jump from Eq. \eqref{eq:dg_jump}, where the intensity from the adjacent element is replaced by the boundary condition. For example, the outer boundary condition $I_{\text{out}}$ can then be written as
\begin{equation}
    \mathcal F^{k}_{ij} = \int_{\partial \D^k} \hat{\boldsymbol n}_r \cdot (f)^*\!\left(I_h^k(\ell_i^k),I_{\text{out}}\right) \, \ell_i^k d\x \quad 
\x \in 
\Gamma_{\text{out}}^- \ .
\end{equation}
This, however, means that the boundary condition is only satisfied in an average sense because it is introduced as a variational formulation.

The second possibility is to directly manipulate the global matrix $\mathcal A$ and the load vector $\mathcal B$. For any degree of freedom $\alpha$, 
located at an element's face where a boundary condition needs to be applied, all entries $\mathcal A_{\alpha \beta, \alpha \neq \beta}$ are set to 
zero, while $\mathcal A_{\alpha \alpha}$ is set to unity. The $\alpha$-th component of the load vector is then set to the respective boundary 
condition. This ensures that the boundary conditions are always met exactly. For our implementation, we chose this second variant.

\subsection{Solving the system of equations}

GetFEM++ itself provides some basic iterative, numerical methods to solve the system of equations \eqref{eq:problem_lse} optimised for sparse 
matrices, including a small set of preconditioners. In this study, we use the GMRES method with an ILU preconditioner. 

However, if required, the sparse matrix $\mathcal A$ and the load vector $\mathcal B$ can also be forwarded to an external solver, such 
as PETSc or MUMPS, a parallel sparse direct solver library, for example.

\end{appendix}

\end{document}